\documentclass[
    aps, prl, reprint,
    superscriptaddress,
    amsmath, amssymb, amsfonts,
]{revtex4-2}
\usepackage{graphicx}
\usepackage{dcolumn}
\usepackage{bm}

\usepackage{booktabs}
\usepackage{siunitx}

\begin{document}
\title{Fluctuation-Response Relations and Memory in Tissue Cell Number Dynamics}
\author{Johannes\,C.\,Krämer}
\author{Jens Elgeti}
\email{j.elgeti@fz-juelich.de}
\affiliation{
Theoretical Physics of Living Matter, Institute for Advanced Simulation, Forschungszentrum Jülich, D-52425 Jülich, Germany}
\date{\today}
\begin{abstract}
    The longevity of multicellular systems requires finely tuned feedback mechanisms to control the balance of cellular proliferation and loss.
    To understand the microscopic details governing feedback during homeostasis, it is crucial to quantify the feedback using a robust, minimally invasive method.
    Existing methods typically depend on perturbing the system to quantify the response.
    Such perturbations may alter cell physiology and are often technically challenging to implement.
    We propose a steady-state fluctuation-analysis framework to measure the dynamic response coefficient of cellular tissues.
    Under the minimal assumption of crowding feedback and Markovian noise, we derive a corresponding theoretical description.
    We validate this framework against three models of increasing complexity: a simple stochastic simulation, an agent-based mechanical model of tissue growth, and a multi-scale model of membrane-potential-regulated growth.
    These tests reveal the suitability of using passive fluctuations to characterize the tissue response parameters. 
    Furthermore, this approach can help discover novel phenomena; in particular, we uncover memory effects arising from the non-equilibrium nature of growth.
    This work opens new avenues to quantify growth parameters in living tissues.
\end{abstract}
\maketitle

\section{Introduction}
Internal energy consumption and environmental noise continuously drive biological systems out of equilibrium~\cite{Ramaswamy2010,Marchetti2013}.
To maintain a stable steady state despite a highly dynamic environment, a robust feedback mechanism is required.
Yet, this feedback is not easily accessible, and existing measurement protocols are not only complicated but often invasive.
Moreover, quantifying possible changes in the dynamic response to perturbations first requires a method to quantify the homeostatic response mode.
It is important to develop methods, that are minimally invasive, ideally passive, and still provide quantitatively correct estimates of the response mode.

In thermal equilibrium, the amplitude of fluctuations and the linear response to perturbation are related by the fluctuation-dissipation theorem~\cite{Callen1951,Kubo1966}.
Active processes, which internally consume energy, break this relation and can cause differences between active and passive measurement protocols~\cite{Turlier2016}.
Yet, already purely fluctuation-based methods, which connect the amplitude of fluctuations with response, can provide insight into the dynamics~\cite{Sato2003}.
In this context, correlation functions are a valuable tool to access information in fluctuating systems~\cite{Kirkpatrick2025,Knotz2024,Muenker2024}.
Recently, many statistical methods for confined random walkers have been derived~\cite{Cherstvy2018,Meyer2021,Meyer2023}.
However, these models are typically derived for ideal systems, and applying them to the complexity of real-world biological systems requires additional effort.

The response of cellular tissues is commonly studied by applying external stresses, i.\,e.\,perturbing the cells.
In vitro this can be done with stretching devices, revealing biomechanical properties of cells~\cite{Constantinou2023,Bighi2024}.
To make these devices compatible with live-cell imaging and realize stretching over long periods is challenging.
Cyclic stretching devices aim to overcome the limitation of limited stretch duration and in addition probe the mechanoresponse to oscillatory cues, often sensed in the natural environment of cells~\cite{Poellen2023}.
Further, realistically engineering the in vivo micro-environment poses difficulties~\cite{Bertoncello2019}.
In vivo stretching has been recently introduced to overcome limitations of in vitro models~\cite{Aragona2020}. However, such methods require surgery, which limits their non-invasiveness.

Here, we extend fluctuation-based response measurements to cellular tissues.
First, we derive the population dynamics of a self-renewing cell population with density feedback on the division rate, under the assumption of stochastic apoptosis and division processes close to the homeostatic state.
We find that the resulting model can be approximately mapped onto a Brownian particle confined by a harmonic potential. 
This system is a well-studied, for example, in the context of beads trapped in optical tweezers.
This allows us to apply a number of statistical methods derived earlier to quantify the tissue response coefficient from steady-state cell number trajectories.
The method is illustrated on data generated from a coarse-grained numerical solution of a discrete-time stochastic birth-death (SBD) process describing the population dynamics, where the ground truth of the response is known.
Further, we test two agent-based simulation models of increasing complexity, and compare results against perturbation protocols.
By analyzing deviations between theory and simulation data, we show how our method can be used to reveal differences in the underlying microscopic mechanisms.
We present how advanced theories, like a relaxing memory of the division rate caused by non-Markovian processes, can be identified and resolved within our framework.

Our novel approach to quantitatively measure the systems response relies purely on passive measurements of cell number trajectories, allowing for straightforward, non-invasive implementation.

\section{Fluctuation response in systems with crowding feedback}
In the absence of noise, we write the evolution of a cellular system consisting of $N$ cells of the same cell type is given by
\begin{equation}
    \partial_t N = (k_d-k_a) N = \Delta k N ,
    \label{eq:PDEcellnumbers}
\end{equation}
where we denote the net bulk growth rate $\Delta k = k_d - k_a$, representing the difference of proliferation via division with rate $k_d$ and cell loss, e.\,g.\,through apoptosis, with rate $k_a$.
For $\Delta k=0$, the number of cells remains constant, and this state is referred to as \textit{homeostatic state}.
We define the cell turnover rate $k_+=(k_d+k_a)/2$.
At homeostasis $k_+=k_d=k_a$, which measures the rate at which cells are replaced.
However, division and loss are random events, so that $k_d$ and $k_a$ represent per-cell rates of a stochastic birth-death process.
Even for zero net growth stochastic drift can lead to over- or underpopulation, where the latter may result in extinction. 
Thus, some form of density-feedback is generally assumed \cite{Shraiman2005,Hallatschek2023}.
Recent research indicates that mechano-chemical coupling plays a relevant role for the details of the feedback mechanism~\cite{Hannezo2019,Alasaadi2024R,Ferrai2024}.
Yet, minimal physics-based approaches that assume a simple mechanical relation between division rate and cell density and pressure have been successfully applied to study cellular tissues~\cite{Basan2009,Montel2011,Podewitz2015}. 
Recent work demonstrates how density couples to pressure and membrane potential, which in turn regulates growth via the \textit{HIPPO} pathway~\cite{Mukherjee2026}.
To first order around the homeostatic state, density-based feedback is given by
\begin{equation}
    \Delta k = \varkappa \frac{ \rho^H - \rho}{\rho^H}, \label{eq:ResponseDensity}
\end{equation}
with cell density $\rho = N/V$, $\rho^H$ the homeostatic density, and $\varkappa$ the response coefficient. 
Combining Eq.\,\ref{eq:ResponseDensity} and Eq.\,\ref{eq:PDEcellnumbers} yields the logistic equation, which formally integrates to
\begin{equation}
    \rho(t) = \frac{\rho^H}{1 + e^{-\varkappa (t-t_0)}},
    \label{eq:NoiseFreeRelaxation}
\end{equation}
where $t_0$ is chosen to satisfy the boundary condition $\rho(t=0)=\frac{\rho^H}{1+e^{\varkappa t_0}}$.
This deterministic approach ignores noise and assumes Eq.\,\ref{eq:ResponseDensity} holds for large density variations. 
It is thus ill suited to reliably obtain the response coefficient.

In order to derive a method, which allows us to measure the response in, or close to, homeostasis, we consistently linearize around homeostasis and include noise, leveraging a generalized fluctuation-dissipation approach. 
Assuming Markovian division and apoptosis processes (see SI~\ref{sec:SI:TheoryDetails} for details), we obtain 
\begin{equation}
    \partial_t \Delta\rho = -\varkappa \Delta\rho  + \sqrt{\frac{2 \rho^H k_+}{V}} \eta(t),
\label{eq:PDEdensityfluctuations}
\end{equation}
with $\Delta\rho=(\rho-\rho^H)$, and $\eta(t)$ delta-correlated white noise.
Note that Eq.\,\ref{eq:PDEdensityfluctuations} is identical to the overdamped Langevin equation of an \textit{equilibrium} Brownian particle confined in a harmonic potential with 
 particle position replaced by cell number density, 
and the spring constant by the division feedback response coefficient $\varkappa$.
Noise is not introduced by a finite temperature, but originates from cell turnover, corresponding to a cell density diffusion coefficient $D = \frac{\rho^H k_+}{V}$, which was also found in a different framework~\cite{Ranft2010}.

We have thus derived a reduced linear Markovian approximation that enables us to apply known results for Ornstein-Uhlenbeck (OU) processes, like the stationary distribution, autocorrelation, and power spectral density~\cite{BergSoerensen2004,Cherstvy2018,Meyer2021,Meyer2023}.
In the remainder of this paper, we will use this framework to obtain effective tissue parameters ($\rho^H, k_+, \varkappa$)  from cell number time series data, and to provide additional dynamical quantifiers that reveal tissue behavior beyond this minimal model. 

\paragraph{Homeostatic and mean density}
The \textit{homeostatic density} $\rho^H$ can be directly measured from time average of the cell number trajectory as
\begin{equation}
    \rho^H \equiv \langle \rho \rangle = \frac{1}{M}\sum_{i=0}^{M-1} \rho(t_i),
\end{equation}
where $M$ depicts the number of measurement points.
Due to the symmetry of Eq.\,\ref{eq:PDEdensityfluctuations} around the homeostatic density, the time average correctly estimates the homeostatic mean.
With increasing fluctuation amplitude nonlinearities can play a role (see SI~\ref{sec:SI:DensProbDensFunc}), causing the true homeostatic density to deviate from the time-averaged mean.
Since the relative noise scales with $V^{-1/2}$, this deviation can be minimized by choosing larger measurement volumes.

\paragraph{Mean squared shift and variance}
The cell turnover rate relates to the diffusion coefficient, and the response coefficient to the harmonic confinement strength. Thus, these parameters can be extracted from the short-time diffusion and the variance, respectively.
In practice, measuring the \textit{mean squared shift} ($MSS)$, which we define in analogy of the \textit{mean squared displacement}, frequently studied to describe dynamics of particle systems, as the $MSD$ in cell density, is more reliable.
The $MSS$ is given by
\begin{align}
    MSS(\Delta t) = 
    \langle (\Delta \rho (\Delta t))^2 \rangle 
    &= \langle (\rho(t + \Delta t) - \rho(t))^2 \rangle 
    \label{eq:MSD:measurement}
    \\
    &= \frac{2D}{\varkappa} \left( 1-e^{-\varkappa \cdot \Delta t} \right),
    \label{eq:MSD:theory}
\end{align}
where $\langle \cdot \rangle$ corresponds to time and/or ensemble averages.
(See Ref.~\cite{Jeon2012} for deriving Eq.\,\ref{eq:MSD:theory} for a Brownian particle in confinement).
At small time differences, cell-turnover noise dominates and the $MSS$ exhibits linear behavior ($\lim_{\Delta t \rightarrow 0} MSS(\Delta t) = 2 D \cdot \Delta t$).
At long time lags, feedback constrains the fluctuations and the MSS reaches a plateau equal to twice the variance $\sigma^2$, i.\,e.\,$\lim_{\Delta t \rightarrow \infty} MSS(\Delta t)=2D/\varkappa = 2\sigma^2$ (see SI~Eqs.\,\ref{eq:app:Equivalence_MSS-densfluctcorr}+\ref{eq:app:Variance}).
Thus, measuring the $MSS$, or alternatively the short-time diffusion $D$ and variance $\sigma^2$, provides us with estimates of the remaining tissue parameters: cell turnover rate $k_{+,D}= V D/\rho^H$ and response rate $\varkappa_{\sigma^2}=D/\sigma^2$. 

\paragraph{Density Fluctuation Autocorrelation Function}
An alternative dynamic quantifier is the density fluctuation autocorrelation function, given by
\begin{align}
    C_{\Delta \rho}(\Delta t) &= 
    \langle
    (\rho(t + \Delta t) - \rho^H)
    (\rho(t) - \rho^H)
    \rangle / \sigma^2
    \label{eq:C:measurement}
    \\
    &= e^{-\varkappa \cdot \Delta t},
    \label{eq:C:theory}
\end{align}
where the relaxation time $\tau = 1/\varkappa$ is determined by the response coefficient.
This autocorrelation is equivalent to the $MSS$ (see SI~Eq.\,\ref{eq:app:Equivalence_MSS-densfluctcorr}).
For an infinitely long, time-invariant process, either quantity allows one to determine the response rate $\varkappa$.
However, for finite-length experimental trajectories, or when microscopic details of cellular processes, like division memory effects, deviate from the assumptions made in the linear model, differences can arise.

\paragraph{Power Spectral Density}
Similar to the density fluctuation autocorrelation, the \textit{power spectral density} ($PSD$) is equivalent to the other dynamic quantities presented above.
Indeed, the $PSD$ is defined as the Fourier transform of the autocorrelation function and is thus directly connected to the $MSS$ (see SI~Eq.\,\ref{eq:Def_FourierTransform_PSD_C}).
However, it can provide additional insight into deviations, and its frequency domain representation emphasizes different dynamical scales.
As derived for the analogous Brownian particle system~\cite{BergSoerensen2004}, the $PSD$ is given by 
\begin{align}
    PSD(f) &= \frac{1}{T_{meas}} \left|
        \int_{-T_{meas}/2}^{T_{meas}/2} dt \Delta\rho(t) e^{i 2\pi f t} 
    \right|^2
    \label{eq:PSD_measurement}
    \\
    &= \frac{D/2\pi^2}{f_c^2 + f^2}
    \label{eq:PSD_Theory}    
\end{align}
with corner frequency $f_c = \varkappa  / 2\pi$ and discrete (Fourier) frequency $f = q/T_{meas}$, with $q\in \mathbb{N}$ and $T_{meas}$ the measurement time.

\paragraph{Probability density}
By analogy to the Brownian particle in a harmonic trap, the probability density of observing a certain cell-number density is obtained from the effective Boltzmann-like stationary density for the mapped OU process, given by
\begin{equation}
    p_{eq}(\rho) = 
    \sqrt{\frac{\varkappa}{2 \pi D}} 
    \exp\left(-\frac{\varkappa}{2D} \left(\Delta \rho\right)^2 \right).
    \label{eq:BoltzmannPD}
\end{equation}

Thus, the probability distribution of density fluctuations is Gaussian, centered at the homeostatic density with variance $\sigma^2 = D/\varkappa$.
The Gaussian approximation holds for sufficiently large response rate and system sizes.
For reduced system size or diminished response coefficient, non-linearity  becomes relevant, leading to non-Gaussian distributions (see SI~\ref{sec:SI:nonlinearPotential} for details) and an increased risk of population extinction~\cite{Ai2003}.
Deviations from Gaussianity can be quantified by the skewness $M_3=\left< \left( (\rho - \rho^H) / \sigma \right)^3 \right>$ and the excess kurtosis $\tilde{M}_4 = \left< \left( (\rho - \rho^H) / \sigma \right)^4 \right>-3$.
This static analysis complements the dynamic measures discussed above.\\

We thus establish a simple procedure to estimate tissue parameters and analyze tissue behavior from passive cell-number measurements alone.
In the remainder of this work, we analyze how three different models of increasing complexity behave in this framework, and how a decaying memory kernel can explain deviations.

\section{Application to growing tissue systems}
To test our theory, we investigate three different systems of increasing complexity.
First, we implement a stochastic birth-death process that matches the stochastic differential equation (SDE) described by Eqs.\,\ref{eq:PDEcellnumbers}\&\ref{eq:ResponseDensity}, integrating over the cell number trajectory numerically.
For this system, all input parameters are known, allowing us to evaluate our measurements against ground truth.
The second test case uses data from more complex agent-based simulations. We adapt the two-particle growth (2PG) model~\cite{Basan2011dpd,Podewitz2015} (see SI~\ref{sec:SI:SimulationModels:2PG}) to simulate a growing cell colony in two dimensions.
Intercellular mechanical interactions are modeled explicitly and implicitly control the division rate, while apoptosis is modeled as a stochastic process with fixed rate $k_a$.
Finally, we turn to a sophisticated mechano-electro-osmotic model (see SI~\ref{sec:SI:SimulationModels:VMPG}).
Cells are represented by a Voronoi tessellation, taking into account more complex shape effects.
In addition to mechanical forces, membrane potential and internal ion concentration are crucial for cell cycle dynamics.
In this Voronoi cell membrane potential growth (VMPG) model, coarse-grained tissue parameters are not known \textit{a priori} and due to the complexity of the simulation, system sizes are limited.

\subsection{Stochastic Birth-Death (SBD) Process}
\begin{figure*}[htbp]
    \centering
    \includegraphics[width=0.9\linewidth]{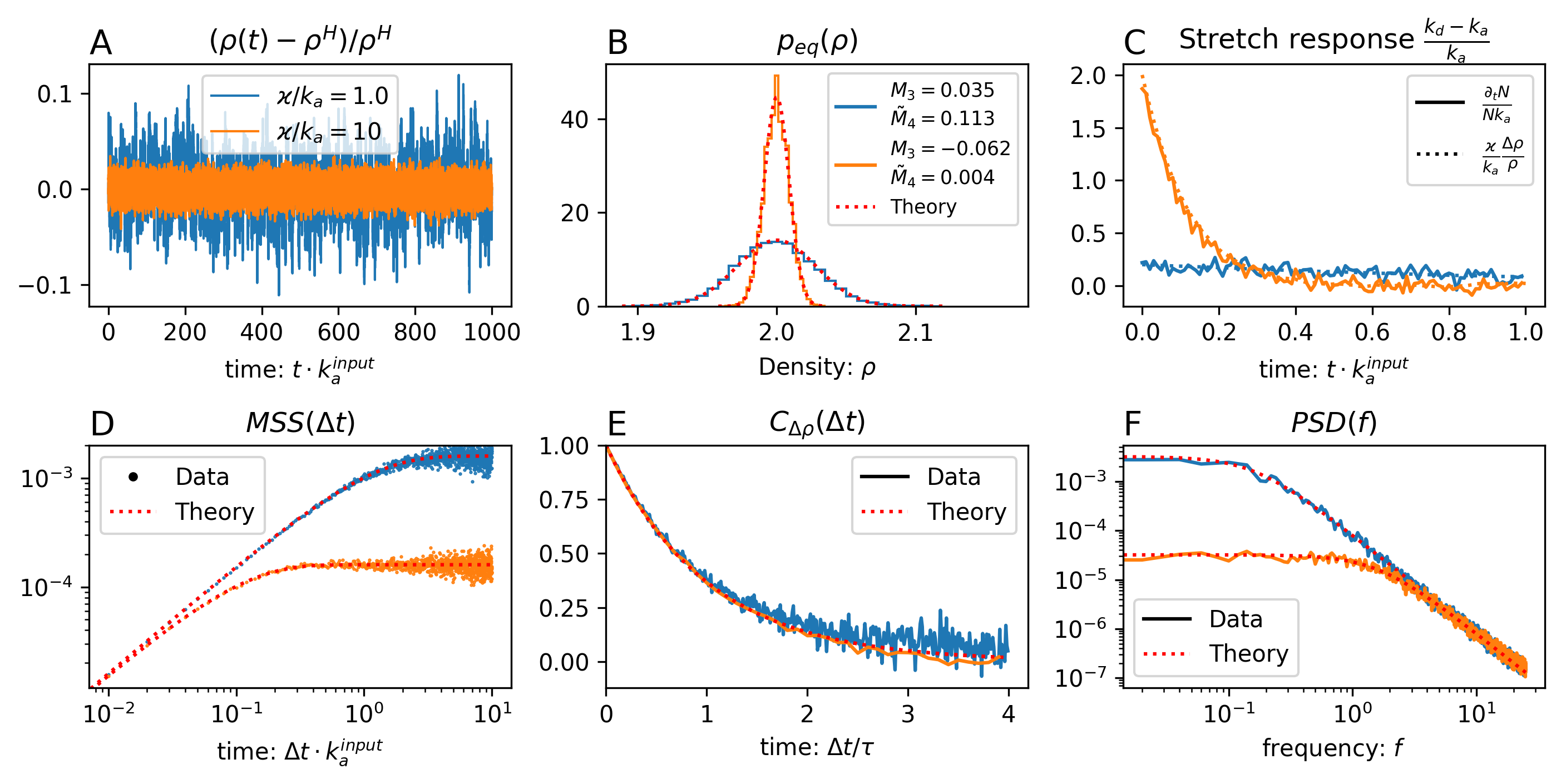}
    \caption[Analysis of the stochastic model trajectories against ground truth parameters.]{
        \textbf{Analysis of the stochastic model trajectories against ground truth parameters.}
        (A) Steady-state trajectories of the SBD model for two response rates $\varkappa$.
        (B) Density probability density $p_{eq}(\rho)$ as function of $\Delta \rho = \rho - \rho^H$, skewness $M_3$ and excess kurtosis $\tilde{M}_4=M_4-3$, where $M_4$ is the kurtosis, are given in legend.
        (C) Net growth rate, measured via Eq.\,\ref{eq:PDEcellnumbers} (solid lines) and Eq.\,\ref{eq:ResponseDensity} (dotted lines), following 20\,\% variation of cell density, averaged over 20 realizations.
        (D) Mean squared shift ($MSS$) of density as a function of time lag
        and (E) density fluctuation autocorrelation function as a function of time lag rescaled by relaxation time.
        (F) Power spectral density ($PSD$) as function of frequency $f$.
        For all figures, two different values of $\varkappa/k_a = 1$ and $10$ are shown in blue and orange, respectively.
        Simulations were performed with $N^H=5000$ cells and for target density $\rho^H=2$, and statistics are aggregated over $1000$ generations.
    }
    \label{fig:Fig1_SBD_example}
\end{figure*}

We implement a stochastic process that closely represents the model~(Eqs.\,\ref{eq:PDEcellnumbers}\&\ref{eq:ResponseDensity}) with well-defined tissue parameters.
A population of $N$ cells in a volume $V$ evolves over time in discrete time steps $\delta t$.
Cells are lost via a Poisson process with constant apoptosis rate $k_a=k_+(\rho^H)$, while the division rate adapts according to the crowding feedback in Eq.\,\ref{eq:ResponseDensity}.
We integrate the stochastic model for a wide variety of apoptosis rates, response coefficients, homeostatic densities, and system sizes, 
but exclude parameters for which density fluctuations are too large ($3\sigma \gtrsim \rho^H$).
The simulations agree well with our theoretical predictions.

Figure~\ref{fig:Fig1_SBD_example} illustrates the behavior for a finite number of cells and demonstrates the match between simulation and theory for two response coefficients ($\varkappa/k_a=1$ and $10$).
It shows the steady-state cell density trajectories, probability density $p_{eq}(\rho)$, division rate response to a 20\,\% volume increase (stretch), $MSS$, autocorrelation function $C_{\Delta \rho}$, and $PSD$.
Analytic predictions using the input parameters of the simulation are shown together with numerical data.
As expected, the data closely follow the analytic predictions, confirming our derivations.

Besides confirming our calculations, these simulations allow us to assess the accuracy of parameter estimation from finite data.
Naturally, tissue parameters can be estimated by fitting the described quantifiers, however limited data lead to limited accuracy.
The two characteristic time scales of the system, i.\,e.\,apoptosis and response rates, dictate the data requirements.
Frame rate must be fast enough to resolve the noise contribution, ensuring the linear regime of the short-term $MSS$.
Trajectories must be long enough to cover the full decorrelation time $\tau$ and reach the $MSS$ plateau.
Furthermore, sufficient statistics are required, as variance estimation is sensitive to errors in the mean density measurement.

\subsection{Agent-based tissue simulations with mechanical feedback--2PG}
\begin{figure*}[htbp]
    \centering
    \includegraphics[width=.9\linewidth]{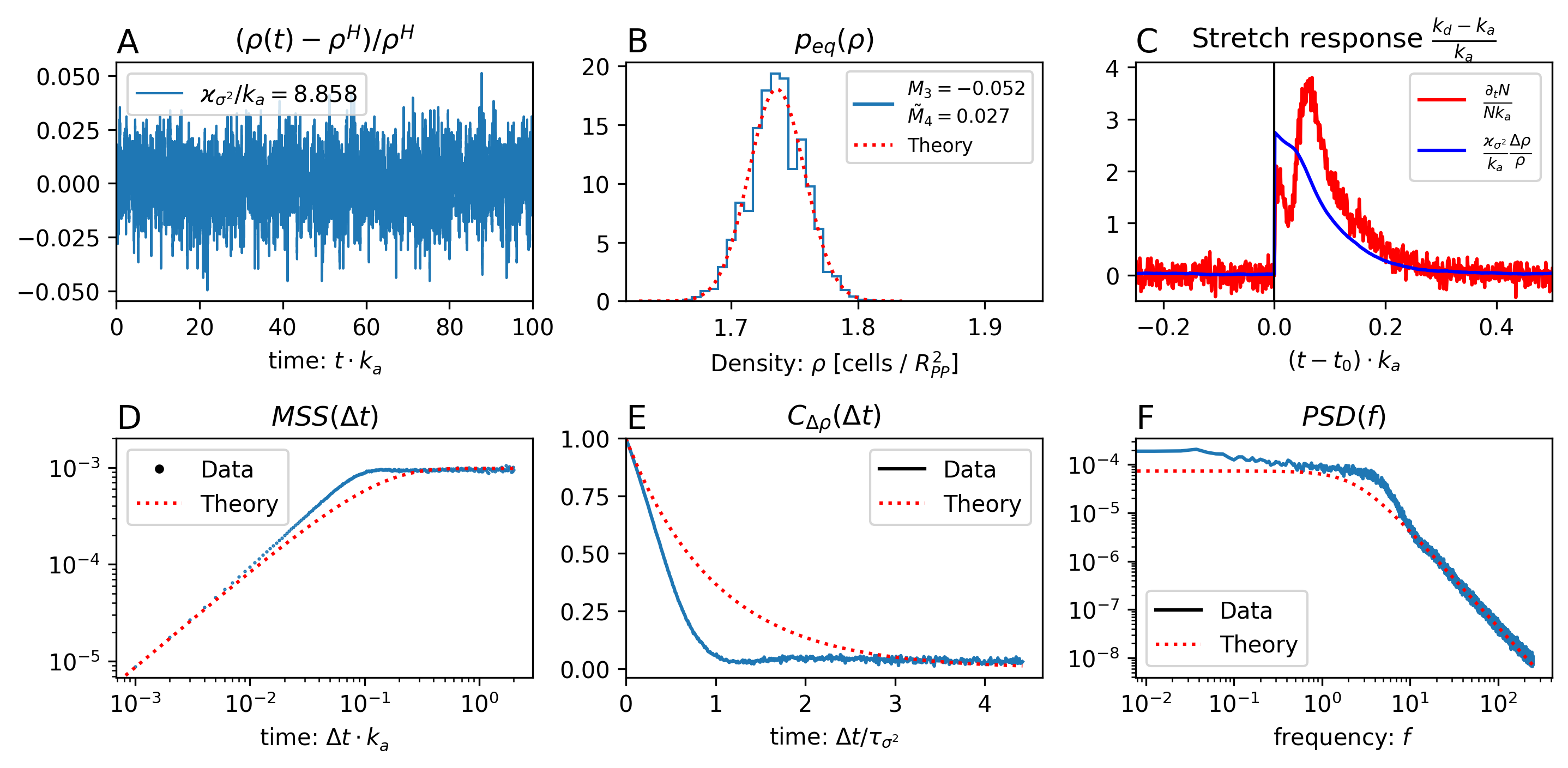}
    \caption{
        \textbf{Fluctuation-response in agent-based simulations}
        (A-F) Same as Fig.\ref{fig:Fig1_SBD_example}, but measured in the 2PG model, where $\varkappa_{\sigma^2}$ was determined by measuring the variance of density fluctuations and inserted in the same theory equations.
        (C) Perturbation is realized by stretching the substrate, such that cell density is reduced by $\sim 20\%$.
        Following the stretch, division rate is measured directly from the current cell number change (Eq.\,\ref{eq:PDEcellnumbers}, red line) and as function of density variation (Eq.\,\ref{eq:ResponseDensity}, blue line), result is averaged over 20 realizations.
        Simulations used for statistical analysis (B, D-F) ran for 4000 cell generations with an area of $(20R_{PP})^2$, where $R_{PP}$ is the cell interaction radius. The resulting homeostatic cell density is $\rho^H = 1.74 \text{cells}/ R_{PP}^2$.
    }
    \label{fig:Fig3_2PGsimu_fluctuationresponse}
\end{figure*}
Coarse-grained models with known input parameters are a valuable for illustrating our methodology.
However, real systems come with additional layers of complexity that may influence behavior.
In simulations of the two particle growth (2PG) model, cells are represented by two point particles.
While apoptosis occurs at a fixed rate, division is driven by mechanics. 
The two particles that represent a biological cell repel each other via an active, repulsive growth force and once a critical separation is reached, the cell divides. 
Combined with mechanical interactions, including excluded volume, attraction, and dissipation, this growth force couples cell division and growth response to the mechanical environment, resulting in a division-apoptosis balance ($k_d = k_a$) at homeostasis.
Yet, how cells respond exactly to density fluctuations is not known a priori.
Simulations are sufficiently fast, that very long trajectories can be simulated for moderately sized systems (see Fig.\,\ref{fig:Fig3_2PGsimu_fluctuationresponse}), yielding robust observables .

Obtaining the turnover rate via the short term diffusion yields excellent agreement with the known ground truth of the set apoptosis rate $k_a$.
The density distribution is nearly Gaussian and symmetric, with skewness $M_3=-0.052$ and excess kurtosis $\tilde{M}_4=0.027$ (see Fig.\,\ref{fig:Fig3_2PGsimu_fluctuationresponse}\,B).
This confirms that the average can be identified as the homeostatic density.
While $\varkappa_{\sigma^2}$ is well-defined, verifying whether it corresponds to the true response coefficient $\varkappa$ requires independent validation.

\begin{figure*}[htbp]
    \centering
    \includegraphics[width=.9\linewidth]{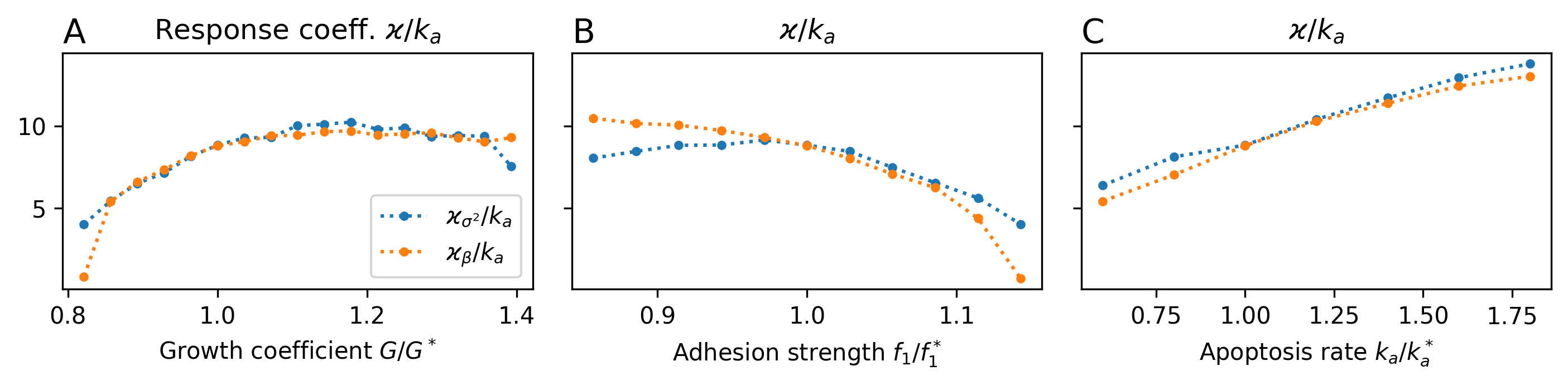}
    \caption{
        \textbf{Comparison of fluctuation-response and perturbation response parameters.}
        For variations of 
        (A) cell growth coefficient $G$,
        (B) cell adhesion strength $f_1$, and
        (C) apoptosis rate $k_a$
        the response coefficients $\varkappa_{\sigma^2}$ and $\varkappa_\beta$ are measured via steady-state fluctuations and volume rescaling, respectively.
        Parameters are scaled by base parameters $G^*=28.0$, $f_1^*=7.0$, and $k_a^*=0.01$.
    }
    \label{fig:Fig5_2PGsimu_comparisonKappa}
\end{figure*}
A conceptually orthogonal approach is to measure the response coefficient via external perturbations.
In simulations, density can be modulated by expanding or compressing volume.
We use a constant volume growth ensemble where the volume is rescaled with a small, constant rate $\beta$, and cell positions are adjusted to realize homogeneous density variations~\cite{Podewitz2015} (see SI~\ref{sec:SI:SimulationModels:2PG:compexp} for details).
The change in division rate and density is measured as a function of the rescaling rate $\beta$.
Fitting Eq.\,\ref{eq:ResponseDensity} to data allows us to measure the perturbation response coefficient $\varkappa_\beta$ (see SI~Fig.\,\ref{fig:app:Fig4_2PGsimu_compressionexpansionresponse}).
This perturbation protocol is numerically and analytically complex, highlighting the simplicity of our fluctuation-based approach. 
Both methods (variance and perturbation) give very similar results over a range of simulation parameters as shown in Fig.\,\ref{fig:Fig5_2PGsimu_comparisonKappa}.

We have varied the growth coefficient $G$, which controls the active growth force of cells, intercellular adhesion strength $f_1$, and apoptosis rate $k_a$.
The parameter scan hints increasing deviations, when parameters are tuned far from the base parameter set.
For high $k_a$ or $f_1$ or low $G$, tissues in simulation may end in a partially teared state, losing confluency.
Conversely, cells get jammed or crystallization effects arise, and their dynamics are again altered.
For a wide range of parameters, however, we find good agreement of both methods.

Again, this confirms the validity of estimating tissue parameters via the cell number and its fluctuations.
While parameters can be obtained via perturbation methods, the fluctuation analysis is less intricate and non-invasive.

All dynamic quantifiers ($MSS(\Delta t)$, $C_{\Delta \rho}$, $PSD(f)$, see Fig.\,\ref{fig:Fig3_2PGsimu_fluctuationresponse}\,D-F),
however, suggest that indeed the particle-based simulations display a more complex response.
A significant deviation at $0.1/k_+$, i.\,e.\,time scales significantly shorter than the relaxation time $\tau=1/\varkappa$~(Eq.\,\ref{eq:C:theory}), is found in all quantifiers.
The autocorrelation shows best how that accelerated relaxation leads to full decorrelation before $t=\tau_{\sigma^2}$.
Importantly, the steady-state fluctuations not only characterize long-term response but even reveal details of the underlying microscopic cellular model.

We speculate, that this deviation comes from a form of memory in the particle-based simulations:
Cell division is not an instantaneous random process, but results from cells growing against the surrounding with a finite force.
They only divide, when a critical size is reached.
Therefore, division rate increase after density is lower \textit{after some time}!
Figure~\ref{fig:Fig3_2PGsimu_fluctuationresponse}\,(C) shows the net growth rate $k=\partial_t N / N$ and the expected growth rate $\varkappa_{\sigma^2} \Delta \rho / \rho$ after a 20\% stretch.
Indeed, cell number does not respond immediately, showing a short delay.
We investigate this effect  in more detail further below.

\subsection{Voronoi model with membrane potential dependent growth--VMPG simulations}
\begin{figure*}[htbp]
    \centering
    \includegraphics[width=.9\linewidth]{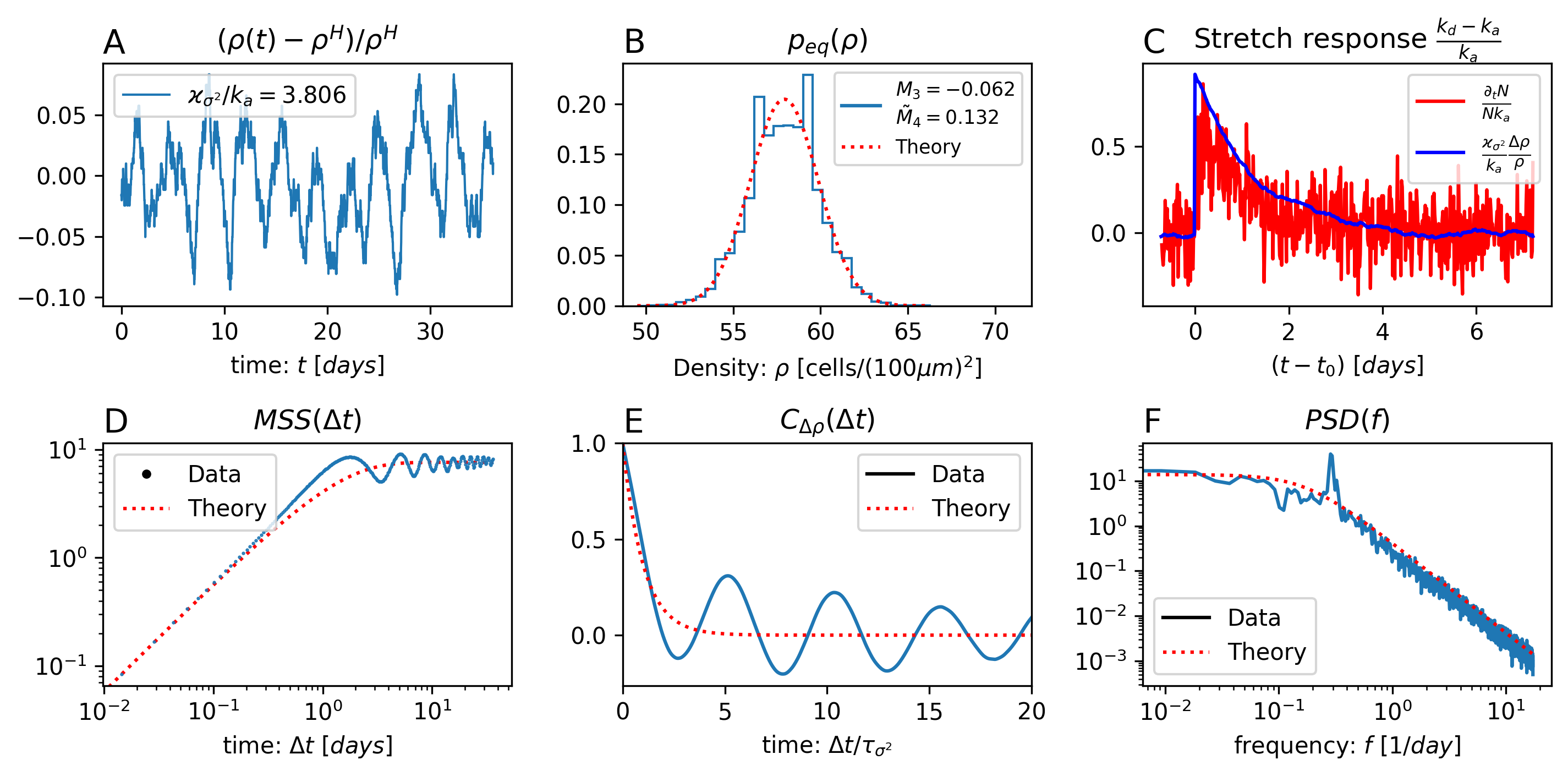}
    \caption{
    \textbf{Membrane-potential-controlled cell cycle dynamics}
    Same as Fig.\,\ref{fig:Fig1_SBD_example} but for an electro-osmotic model originally published in \cite{Mukherjee2026}, wild type parameters.
    Response rate is determined from variance.
    The underlying adder model apparently results in some form of synchronization, which results in oscillations of the MSS and autocorrelation function and a peak at a frequency of approximately every third day in the PSD.
    Data of 16 independent runs, each 100 days long, are aggregated in statistical analysis. Stretch response is averaged over 32 independent realizations, system size is $(200\mu m)^2$.
    }
    \label{fig:Fig6_CellPaperResponse}
\end{figure*}

Finally, we turn our attention to a much more detailed model calibrated to fit experimental data.
We analyze cell-number trajectories generated with agent-based simulations of a model published in reference~\cite{Mukherjee2026} (see SI~\ref{sec:SI:SimulationModels:VMPG}).
In this model, none of the tissue parameters are imposed or known a-priori.
Only the apoptosis rate can be assumed to be one in five days, if membrane potential never exceeds $\si{-70}{mV}$, which seems to be the case for the data used.
Consistent with this, the linearity of the short-term diffusion coefficient matches the expected value for apoptosis.
Note that the numerical cost of these simulations prevents simulations of large systems or obtain the response coefficients via perturbation simulations. 
We still obtain reasonable statistics by aggregating data from independent runs.
As for the particle-based simulations, the density distribution is rather symmetric ($M_3 = -0.062$, $\tilde{M}_4=0.132$, see Fig.\,\ref{fig:Fig6_CellPaperResponse}\,B), so that we obtain a homeostatic density $\rho^H=<\rho>=57.92~\text{cells}/(100\mu m)^2$.
The raw cell number trajectory already hints that the system is not just randomly fluctuating in steady state. Regular patterns indicate ongoing oscillations. 
We suspect that these are caused by the underlying \textit{adder}-growth model and may be enhanced by the small scale of the simulations, which can favor synchronization.
This effect is investigated in more detail below.

The periodicity can be assigned to a peak at a specific frequency of roughly every fifth day in the power spectral density (see Fig.\,\ref{fig:Fig6_CellPaperResponse}\,F), matching the apoptosis rate.
It is also reflected in oscillations of the $MSS$ and autocorrelation function (see Fig.\,\ref{fig:Fig6_CellPaperResponse}\,D-E), which shows that oscillations are weakly damped and continue for many days.
Despite the significantly different dynamics, the response of $\varkappa_{\sigma^2}/k_a=3.806$ coefficient determined from the variance still describes the overall relaxation process well.
\subsection{Memory in Division control}
Both agent-based simulation models showed differences between in their dynamic relaxation compared to our theory.
Relaxation seems to be much faster than expected from the long-term response rate, defined by the plateau value of the different dynamic quantifiers, i.\,e.\,the variance of fluctuations.
Also, oscillations hint some synchronization of growth in the VMPG model.
To better understand this difference we investigate the effect of division memory on the fluctuations.

\paragraph{Cellular division with memory effects}
\begin{figure*}[htbp]
    \includegraphics[width=.9\linewidth]{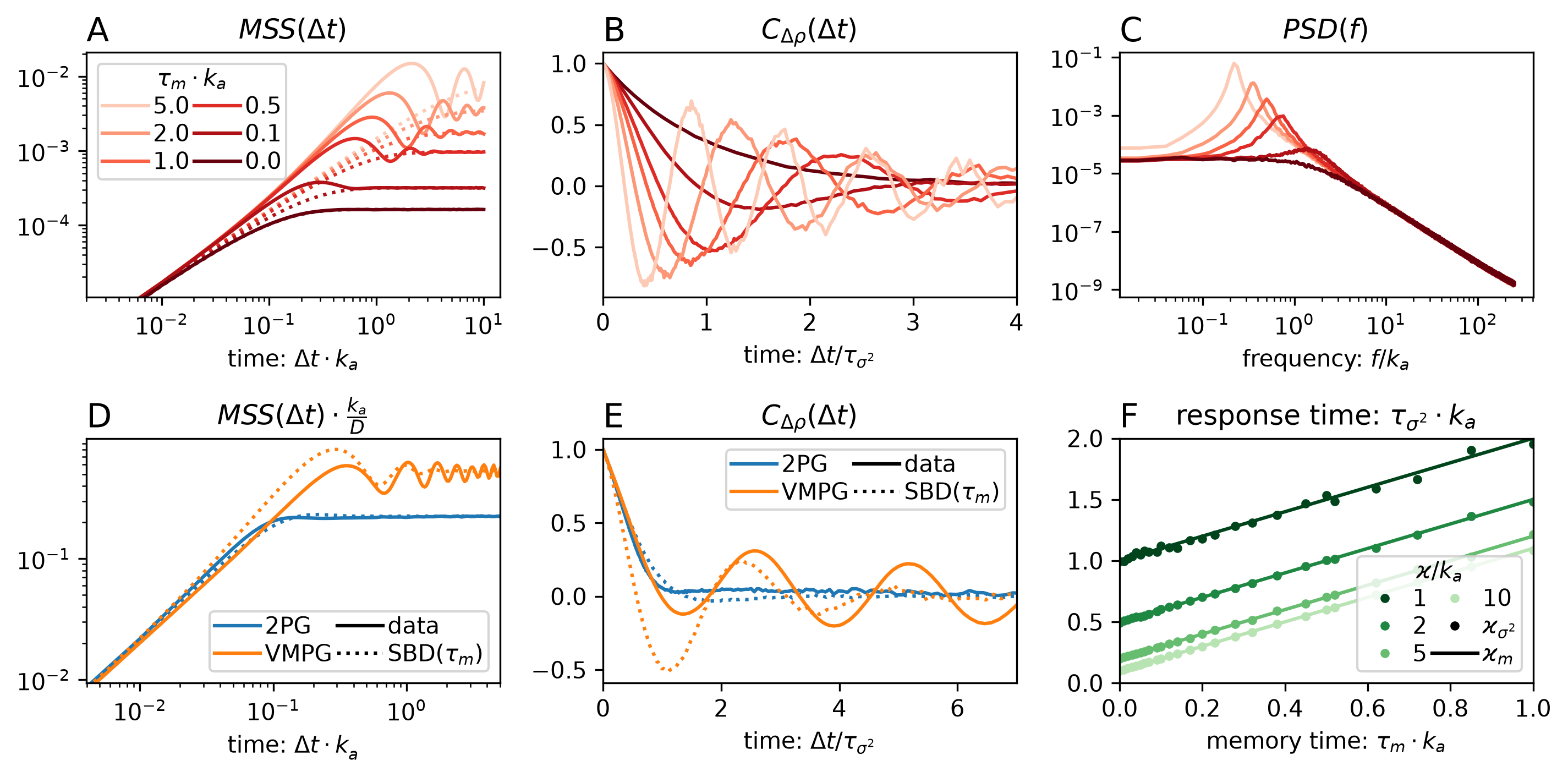}
    \caption{
        \textbf{Division memory effects on fluctuation response.}
        (A) $MSS$, (B) $C_{\Delta\rho}$, and (C) $PSD$ for the SBD process with division memory as described in Eq.\,\ref{eq:DivisionTime-memory-differential}.
        In (A), the memory-free MSS for measured $\varkappa_{\sigma^2}$ (dotted line) and $MSS$ in the presence of memory (solid line) are shown in comparison.
        Color encodes memory relaxation time $\tau_m$ and input response rate is $\varkappa/k_a=10$.
        (D) $MSS$, normalized by $k_a/D$ to align curves,  and (E) $C_{\Delta \rho}$ for 2PG simulations (blue) and VMPG simulations (orange).
        The SBD model with fitted memory time is shown as dotted lines, where $\tau_m \cdot k_a = 0.035$ and $\tau_m \cdot k_a = 0.22$ are used for the 2PG and VMPG model, respectively. 
        For VMPG simulations, long-standing oscillations are not correctly explained by exponentially decaying memory alone.
        Note that $\varkappa$ of SBD simulations is chosen such that resulting response rate matches $\varkappa_{\sigma^2}$ of particle-based simulation.
        (F) Effective relaxation time $\tau_{\sigma^2}=1/\varkappa_{\sigma^2}$ measured from SBD memory simulations (dots) as a function of memory time $\tau_m$, with input response rate $\varkappa$ encoded in color. Solid line represents response time following from superposition of cell-intrinsic response and memory relaxation, $\tau_{\sigma^2} = \tau_m + 1/\varkappa$.
    }
    \label{fig:SBE-division-waitingtime-memory}
\end{figure*}

Instead of immediately increasing the division rate in response to lower density, cells first need to grow.
This (short) growth period is then followed by an excess number of divisions.
To test this idea, we adapt the division rate update rule in the stochastic model.
A generic form of memory is given by integrating over the past states of the system
\begin{equation}
    k_d^m(t) - k_a = 
    \int_{-\infty}^t dt'~\varkappa \frac{\rho^H - \rho(t')}{\rho^H}
    w_m(t-t').
    \label{eq:DivisionTime-memory}
\end{equation}
where $k_d^m(t)$ is the resulting memory-dependent division rate and $w_m(t)$ a time dependent weight function, which describes the memory process.
In general $w_m(t)$ is unknown, so we assume for simplicity that the system loses memory of the past in an exponential relaxation process.
Then, the weight function is given by
\begin{equation}
    w_m(t-t_0) = \frac{e^{-(t-t_0)/\tau_m}}{\tau_m}
\end{equation}
with memory relaxation time $\tau_m$. 
Division by $\tau_m$ ensures, that for $\tau_m\rightarrow 0$ the memory-free division rate, given by Eq.\,\ref{eq:ResponseDensity}, is recovered.
Alternatively, this memory can be written as a dynamic equation for the effective growth rate
\begin{equation}
    \partial_t (k_d^m(t) - k_a) = \frac{1}{\tau_m} \left( 
        \varkappa \frac{\rho^H - \rho(t)}{\rho^H} - (k_d^m(t) - k_a)
    \right).
    \label{eq:DivisionTime-memory-differential}
\end{equation}
that is, the growth rate relaxes on times $\tau_m$ towards the linear term of the memory-free theory.

The chosen memory kernel enables an analytic prediction for the change of the measured system response $\varkappa_{\sigma^2}$ as a function of the cell intrinsic response $\varkappa$ and memory relaxation time $\tau_m$ (see SI~Sec.~\ref{sec:app:memorykernel_analytic}). We find, that for the linearized system, superposition of the cell response relaxation and memory response relaxation describes the system response, i.\,e.: $\tau_{\sigma^2} = 1/\varkappa_{\sigma^2}=1/\varkappa+\tau_m$.
Oscillations of the relaxation process occur for $4\varkappa\tau_m>1$, and are a consequence of the coupling between density of density feedback and growth rate update. 

Implementing this process into the SBD simulations ($SBD(\tau_m)$), we find, that with increasing memory relaxation time $\tau_m$ the same features as observed in agent-based simulations are recovered (see Fig.\,\ref{fig:SBE-division-waitingtime-memory}\,A-C).
First, we investigate the effect of an increasing memory relaxation time $\tau_m$.
All other parameters of the simulation are kept identical.
We find that effectively the measured response coefficient $\varkappa_{\sigma^2}$ decreases.
These results confirm our theory prediction, that it is described by adding up the fluctuation relaxation time and memory relaxation time (see Fig.\,\ref{fig:SBE-division-waitingtime-memory}\,F).
In comparison with the memory-free theory, deviations progressively build up with increasing memory time, resulting in the expected oscillations, displayed in the mean squared shift ($MSS$) and density autocorrelation ($C$) and a corresponding peak in the frequency spectrum.

We fit this adapted theory to our agent-based simulation data (see Fig.\,\ref{fig:SBE-division-waitingtime-memory}\,D-E).
The updated input response coefficient is not a free parameter, but calculated from the previously measured response coefficient and fit value of the memory relaxation time.
For the 2PG model, we find that a relatively short memory time $\tau_m\cdot k_a=0.035$ yields good agreement between the memory SBD model and growth model simulation data.
However, small differences remain.
This suggests that, while a collective memory effect explains the previously observed differences well, higher order effects are adding even more complex features.

The VMPG model cannot be fitted sufficiently well by the exponentially decaying memory kernel.
Even though we can recover oscillations with increasing memory times, frequency and decay rate differ from simulation results.
A memory time of $\tau_m\cdot k_a=0.22$ comes closest to describe the general trend.
A more complex memory kernel function may be required to capture the long-standing oscillations, which will be explored in future work.
For example, one could introduce a weakly damped oscillating function, which is introduced by synchronization events from the underlying adder-growth model.
Overall, it appears that memory plays a key role in describing and understanding the finer details of tissue response to density variation.

\section{Discussion}
Biological systems are constantly driven out of equilibrium by intrinsic or environmental noise.
Steady state must be restored, resulting in fluctuations of the system around the steady-state.
We have shown that these fluctuations can be very useful in quantifying the dynamic response of a noisy system.
Mapping a growing, self-renewing tissue population to an analog system of a Brownian particle confined in a harmonic oscillator allowed us to derive a set of static, i.\,e.\,cell density probability density, and dynamic, i.\,e.\,mean squared shift, autocorrelation function, and power spectral density, which can together quantify the dynamic response coefficient and detect indications to non-equilibrium effects, like finite memory of the division rate.

We illustrate the correctness of our mapping with help of cell number trajectories for a simplified, coarse-grained system, where direct feedback of the division rate on the current density is implemented.
In two agent-based models of increasing complexity, first a mechanistic two-particle growth model and second a mechano-electro-osmotic model, confirm the robustness of the variance-based method to estimate the response coefficient. 
For the 2PG model, the variance response coefficient was compared to a response coefficient measured via compression-expansion modulations of the tissue. Over a large parameter range both methods show quantitative agreement.
In addition, we find that differences between our theory and the simulation data can be resolved by a memory of the division rate.
Due to finite growth periods, the division rate requires a finite time to adapt.
This non-Markovian mechanism can resolve differences between two particle growth model and our theory.
In the more complicated mechano-electro-osmotic model, long-standing oscillations could not be fully resolved by a simple exponentially decaying memory kernel.
More sophisticated division update rules are required to account for the synchronization effects, which we assume to arise due to the underlying adder-mechanism of the growth model.
Here, we have limited the models to mechanically inspired models, but our theory generally holds for other types of density-based feedback, like biochemical quorum sensing.

Recently, other studies did show, that a combination of active and passive rheology~\cite{Turlier2016}, but also three-point correlation functions of passive fluctuations only~\cite{Muenker2024,Knotz2024}, can indicate the non-equilibrium nature of biological systems.
In single-cell experiment, fluctuations of the cell are probed.
While a comparison between active and passive rheology may cause spurious artifacts stemming from the required external force to perturb the system, in particular passive-only methods are of interest.
We did not find any additional signs of activity in the mean back relaxation ($MBR$) function, a three-point correlation function calculated by averaging over the ratio of future and past displacements (see SI~\ref{sec:SI:MBR}).
As expected for Gaussian processes, which we found in all of our systems, the $MBR$ contains no additional information beyond what is encoded in the $MSS$ and can be expressed in terms of the latter.
On the other hand, our results suggest that cellular tissues can serve as an interesting model to investigate non-equilibrium physics.

In this work, we focused on tissues with a cellular tissue consisting of a single cell type.
Real tissues consist of various different cell types, which may have different dynamic response coefficient.
If response coefficients vary greatly, classical perturbation measurements are unlikely to resolve all coefficients equally well.
We expect the variance-based measurement protocol to be suitable to obtain the different response coefficients in multicellular systems, when sufficiently long trajectories are recorded.

Finally, microscopy of biological samples comes with additional challenges. 
On one hand, the same section of cells is rarely imaged repeatedly.
Also, even if the exact same position is imaged, cells randomly move in or out of the image section.
Both effects result in an additional noise contribution.
In order to apply our method to experimental data, one has to understand how spatial noise affects cell number fluctuations.
We expect that for large volumes, this surface term plays a minor role but becomes relevant for limited image sections.

\appendix
\setcounter{section}{19} 
\setcounter{figure}{0}
\renewcommand*\thefigure{\thesection\arabic{figure}}
\setcounter{table}{0}
\renewcommand*\thetable{\thesection\arabic{table}}
\setcounter{equation}{0}
\section{Supplementary Information}\label{sec:SI}

\subsection{Details on density fluctuations due to noise}\label{sec:SI:TheoryDetails}
In order to describe the biological, noisy nature of tissues, we assume that both, division and apoptosis are subject to stochastic noise and can be modeled by Poisson processes.
We assume that the cell-number N is large, so that a sufficient number of events is observed in a small time step $\delta t$. 
Then, we can rewrite the change of cell numbers $N$ in terms of Gaussian distributions $\mathcal{N}$ as
\begin{equation}
\begin{aligned}
    \Delta N &= N(t + \delta t) - N(t)
    \\
    &= \mathcal{N}(k_dN\delta t, k_dN\delta t) - \mathcal{N}(k_aN\delta t, k_aN\delta t)
    \\
    &= (k_d  - k_a) N \delta t + \mathcal{N}(0, k_dN\delta t) - \mathcal{N}(0, k_aN\delta t)
    \\
    &= (k_d  - k_a) N \delta t + \sqrt{(k_d + k_a)N \delta t} \mathcal{N}(0, 1)
    \\
    &= \Delta k N \delta t + \sqrt{2k_+N \delta t} \mathcal{N}(0, 1)
    \label{eq:StochasticCellNumberChange}
\end{aligned}
\end{equation}
Note that we assume that the system is already in steady state ($\Delta k=k_d-k_a\approx 0$).
Also, we do assume Markovian noise with finite variance, which leads to the Gaussian distribution at large cell numbers.
Inserting Eq.\,\ref{eq:ResponseDensity} after division by Volume $V$, gives
\begin{equation}
\begin{aligned}
    \partial_t \rho &= -\varkappa \frac{\rho-\rho^H}{\rho^H} \rho + \sqrt{\frac{2 k_+ \rho^H}{V}} \eta(t)
    \label{eq:app:Langevin-density}
\end{aligned}
\end{equation}
with $\eta(t)$ describing a mean centered $\langle \eta(t) \rangle = 0$, uncorrelated $\langle \eta(t) \eta(t')\rangle = \delta(t-t')$ white noise. 
Note that we already made the assumption that number fluctuations are small and replaced $\rho$ in the noise term by $\rho^H$. 
Consequently, instead of multiplicative noise our stochastic differential equation only contains additive noise, simplifying the calculus significantly and allowing us to exploit existing solutions.
For small fluctuations $y = \frac{\rho-\rho^H}{\rho^H} = \Delta \rho / \rho^H$ and densities close to the homeostatic density, this can further be simplified as
\begin{align}
    \partial_t y &= -\varkappa y (y+ 1) + \sqrt{\frac{2 k_+}{\rho^H V}} \eta
    \label{eq:app:Langevin-nonlinear}
    \\
    &\approx -\varkappa y + \sqrt{\frac{2 k_+ }{\rho^H V}} \eta
    \label{eq:app:Langevin-approx}
\end{align}
Eventually, under the assumption of mechanical feedback on growth and only small fluctuations around the homeostatic state, we have derived the equivalent Langevin equation for a Brownian particle confined in a harmonic potential.
This allows us to provide a number of analytic expressions as discussed in the main text and below.
Note that the different notations in terms of $\Delta \rho$ and $y$ only differ in the definition of the diffusion coefficient.

\paragraph{Non-linear confinement potential}\label{sec:SI:nonlinearPotential}
In Eq.\,\ref{eq:app:Langevin-approx}, we have introduced a simplified description analogue to a Brownian particle confined by a harmonic potential.
To better understand this approximation, we can rewrite Eq.\,\ref{eq:app:Langevin-density} as
\begin{equation}
\begin{aligned}
    \partial_t \rho &= \left[ - \partial_\rho U(\rho)\right] + \sqrt{\frac{2 k_a \rho^H}{V}} \eta(t)
    \label{eq:app:Langevin-density-potential}
\end{aligned}
\end{equation}
with (full) feedback potential 
\begin{equation}
    U(\rho) = \frac{\varkappa}{\rho^H} \left( \frac{\rho^3}{3} - \frac{\rho^2\rho^H}{2} \right) .
    \label{eq:app:FeedbackPotential}
\end{equation}
At $\rho=\rho^H$, a stable local minimum is found.
The fixed point $\rho=0$ is unstable and represents a local maximum, or barrier, to population extinction.
The response rate is equivalent to a spring coefficient in the OU-process and describes the coupling strength of the cell density to its homeostatic value.

\paragraph{Density probability density function}\label{sec:SI:DensProbDensFunc}
In equilibrium, the stationary Fokker-Planck equation
\begin{equation}
    D \frac{d^2 p(x)}{dx^2} = \frac{d}{dx} \left[ -\frac{dU(x)}{dx} p(x) \right]
    \label{eq:app:FokkerPlanck_stationary}
\end{equation}
describes the probability density of a drift diffusion process and is solved by
\begin{equation}
    p(x) = p_0 \exp\left( -\frac{U(x)}{D} \right).
    \label{eq:app:FokkerPlanck_solution_general}
\end{equation}
For a harmonic potential $U(x) = \varkappa x^2 / 2$, the solution is the Boltzmann statistics.
In case of the tissue given by
\begin{equation}
    p_{eq}(\rho) = 
    \sqrt{\frac{\varkappa}{2\pi D}}
    \exp\left( -\frac{\varkappa}{2D} \left(\rho-\rho^H\right)^2 \right).
    \label{eq:app:FokkerPlanck_solution}
\end{equation}
For the non-linearized potential, given in Eq.\,\ref{eq:app:FeedbackPotential}, the stationary distribution can only be treated numerically.
Since negative densities are not realistic, normalization condition $\int_0^\infty d\rho p(\rho) =1$ converges.
For a series of response rates and system sizes, we observe that the probability distribution is close to a Gaussian for sufficiently large response rate ($\varkappa/k_a \geq 1$) and system sizes ($N^H \geq 100$).
Importantly, reduction of $\varkappa$ or $N^H$ results in a stronger skewness of the potential and deviation from the Gaussian.
Also, the extinction probability, measured as $p(\rho=0)$, increases for these cases.

\paragraph{Autocorrelation and Variance} \label{sec:app:paragraph:autocorrelation-variance.}
The power spectral density discussed in the main text represents the Fourier space representation of the fluctuation auto-correlation function $C_X(t)$, defined as
\begin{equation}
    C_X(t) = \langle
        (X(t) - \mu) (X(0) - \mu)
    \rangle,
    \label{eq:Def_Autocorrelation}
\end{equation}
where $X(t)$ depicts the observable of our measurement and $\mu$ its mean value.
Now, one can obtain the autocorrelation or $PSD$ from each other via
\begin{equation}
    \begin{aligned}
        C_X(t) &= \int_{-\infty}^{\infty} P_X(\omega) e^{i \omega t} \frac{d\omega}{2\pi}
        \\
        P_X(\omega) &= \int_{-\infty}^{\infty} C_X(t) e^{-i \omega t} d\omega
    \end{aligned}
    \label{eq:Def_FourierTransform_PSD_C}
\end{equation}
with frequency $\omega = 2\pi f$.

For zero time lag, the auto-correlation equals the variance of $X(t)$
\begin{equation}
    \sigma^2_X \equiv Var(X) = \mathbb{E}\left( (X - \mu)^2 \right).
    \label{eq:Def_Variance}
\end{equation}

Inserting Eq.\,\ref{eq:PSD_Theory} in Eq.\,\ref{eq:Def_FourierTransform_PSD_C} for $t=0$ is analytically tractable and yields
\begin{equation}
\begin{aligned}
    \sigma_\rho^2 
    &= C_\rho (t=0)
    \\
    &= \int_{-\infty}^\infty \frac{
		2 \left(k_+ \cdot \rho^H \right) / V
	}{
		\varkappa^2 + \omega^2
	} \frac{d\omega}{2\pi}
    \\
    &= \frac{k_+ \rho^H}{\varkappa V}
\end{aligned}
\label{eq:VarianceResponse}
\end{equation}
Thus, that one can directly measure the response coefficient via the variance of fluctuations.

\paragraph{Equivalence of MSS and density fluctuation correlation function}\label{sec:SI:MSS-C-relation}
Mean squared shift (MSS) of density and the "pair"-correlation of density fluctuations are equivalent.
First, we note, that the MSS of density and density fluctuations is the same.
Next, we derive the relation connecting density autocorrelations with density fluctuation autocorrelation.
\begin{equation}
\begin{aligned}
    &C_{\Delta\rho}(\Delta) 
    \equiv \langle (\rho(t+\Delta) - \rho^H) (\rho(t) - \rho^H) \rangle / \sigma^2
    \\
    &= \langle \rho(t+\Delta)\rho(t) + (\rho^H)^2  - \rho^H(\rho(t+\Delta) + \rho(t))) \rangle/ \sigma^2
    \\
    &= \left(\langle \rho(t+\Delta)\rho(t) \rangle + (\rho^H)^2  - \rho^H \langle(\rho(t+\Delta) + \rho(t))) \rangle \right) / \sigma^2
    \\
    &= \left( \langle \rho(t+\Delta)\rho(t) \rangle - (\rho^H)^2 \right) / \sigma^2
    \\
    &= C_\rho(\Delta) / \sigma^2 - (\rho^H)^2 / \sigma^2
    \label{eq:app:Equivalence_denscorr-densfluctcorr}
\end{aligned}
\end{equation}
In the last step, we introduced the density autocorrelation function $C_\rho(\Delta) = \langle \rho(t+\Delta)\rho(t) \rangle$.
This expression can be found in the MSS, when expanding the expectation values, and allows us to show the equivalence of MSS and $C_{\Delta \rho}$, given by
\begin{equation}
\begin{aligned}
    MSS &= \langle (\rho(t+\Delta t) -\rho(t))^2 \rangle
    \\
    &= \langle 
        (\rho(t+\Delta t)^2 + \rho(t)^2) 
        -2(\rho(t+\Delta t)\rho(t)) 
    \rangle
    \\
    &= 2\langle \rho(t)^2 \rangle
    - 2 \langle 
        (\rho(t+\Delta t)\rho(t)) 
    \rangle 
    \\
    &= 2 \langle \rho(t)^2 \rangle - 2C_\rho(\Delta t)
    \\
    &= 2 \langle \rho(t)^2 \rangle - 2 (\rho^H)^2 - 2 \sigma^2 C_{\Delta \rho}(\Delta t)
    \\
    &= 2 \sigma^2 (1 - C_{\Delta\rho}(\Delta) ).
    \label{eq:app:Equivalence_MSS-densfluctcorr}
\end{aligned}
\end{equation}
Where we have used, that the variance of density fluctuations is given by
\begin{equation}
\begin{aligned}
    \sigma^2 &= \langle (\rho(t) - \rho^H)^2 \rangle
    \\
    &= \langle \rho(t)^2 + (\rho^H)^2 - 2 \rho(t)\rho^H \rangle
    \\
    &= \langle \rho(t)^2 \rangle - (\rho^H)^2.
    \label{eq:app:Variance}
\end{aligned}
\end{equation}
Eventually, we have shown, that for known variance, one can directly measure the response coefficient via density fluctuation autocorrelations.
Also, the characteristic relaxation time is equivalent to the response rate $\tau = 1/\varkappa$.

\subsubsection{Derivation of memory relaxation time}\label{sec:app:memorykernel_analytic}
For the linearized system, the effective relaxation time in the presence of an exponentially decaying memory kernel can be derived analytically.
The coupled differential equations are
\begin{equation}
\begin{aligned}
    \partial_t \Delta \rho &= g(t) \rho^H + \sqrt{2D} \eta(t)
    \\
    \partial_t g(t) &= \partial_t \left( k_d^m(t) - k_a \right) = \frac{1}{\tau_m} \left(
        -\frac{\varkappa \Delta \rho}{\rho^H} - g(t)
    \right)
\end{aligned}
\end{equation}
Note that we linearize the system in homeostasis, with $\rho(t) = \rho^H + \mathcal{O}(\Delta \rho)$.
Rewriting this as 2D linear system, we get
\begin{equation}
\begin{aligned}
    \partial_t\left(\begin{array}{c}
         \Delta \rho(t) \\
         g(t)
    \end{array}\right)
     &= \left(\begin{array}{cc}
        0 & \rho^H \\
        -\frac{\varkappa}{\tau_m \rho^H} & -\frac{1}{\tau_m}
    \end{array}\right)
    \left(\begin{array}{c}
         \Delta \rho(t) \\
         g(t)
    \end{array}\right)
    \\
    &\quad+ \left(\begin{array}{c}
        \sqrt{2D}\eta(t) \\
        0
    \end{array}\right)
\end{aligned}
\end{equation}
To solve the relaxation of this multivariate Ornstein-Uhlenbeck process, see also Refs.~\cite{Lax1960,Godrche2018}, we define the state vector 
\begin{equation}
    \mathbf{X} = \left(\begin{array}{c}
         \Delta \rho(t) \\
         g(t)
    \end{array}\right),
\end{equation}
and the covariance matrix
\begin{equation}
    \mathbf{\Sigma} = \langle \mathbf{X}\mathbf{X}^T \rangle
    = \left(\begin{array}{cc}
        \sigma_\rho^2 & \sigma_{\rho g} \\
        \sigma_{\rho g} & \sigma_g^2
    \end{array}\right).
\end{equation}
With the drift and diffusion matrices
\begin{equation}
    \mathbf{A} = \left(\begin{array}{cc}
        0 & \rho^H \\
        -\frac{\varkappa}{\tau_m \rho^H} & -\frac{1}{\tau_m}
    \end{array}\right), \quad
    \mathbf{Q} = \left(\begin{array}{cc}
        2D & 0 \\
        0 & 0
    \end{array}\right).
\end{equation}
we can then calculate the variances, and related response coefficients, by solving the Lyapunov equation $\mathbf{A}\mathbf{\Sigma} + \mathbf{\Sigma}\mathbf{A}^T + \mathbf{Q} = \mathbf{0}$.
Solving for the different matrix components, we get
\begin{equation}
        \sigma_{\rho g} = -\frac{D}{\rho^H},\quad
        \sigma_{g}^2 = \frac{\varkappa D}{\left(\rho^H\right)^2},\quad
        \sigma_{\rho}^2 = D\left(\frac{1}{\varkappa} + \tau_m \right),
\end{equation}
where the latter equation is an exact analytic expression for the relaxation time $\tau_{\sigma^2} = 1/\varkappa_{\sigma^2} = \sigma_\rho^2 / D$ in presence of a memory kernel.
\\
Besides the analytic expression for the memory relaxation time, we further find the condition for oscillations.
For the drift matrix $\mathbf{A}$, the eigenvalues can be calculated from the equation $\lambda^2 + \lambda/\tau_m + \varkappa / \tau_m=0$, resulting in oscillations for $4\varkappa\tau_m>1$. 
While the real parts of all eigenvalues are always negative, confirming stability, the drift-diffusion process is irreversible ($\mathbf{AQ}\neq\mathbf{QA}^T$), as one expects for a non-equilibrium steady state.

\subsection{Simulation details and additional results}\label{sec:SI:SimulationModels}
Here, we briefly introduce the three different simulation models used to test our theory in the main text.

\subsubsection{SBD model}\label{sec:SI:SimulationModels:sODE}
Eq.\,\ref{eq:PDEcellnumbers}\&\ref{eq:ResponseDensity} are implemented to numerically solve a discrete-time stochastic birth-death process in \verb|python|.
At each time point of the simulation, the number of deaths and divisions are calculated by binomial sampling, where the probability of success is given by the current division rate, or fixed apoptosis rate, multiplied by the time step $\Delta t$ of the simulation.
Then, the current number of cells is updated followed by an update of the division rate.
These simulations do not require a simulation volume, as the update rule for the division rate can be normalized with a target cell number instead. 
Yet, we chose to set a volume to match the cell density of 2PG and VMPG simulations, when fitting our model and a density of $\rho^H_{SBD}=2$ for independent simulations.

\subsubsection{Two particle growth model}\label{sec:SI:SimulationModels:2PG}
\begin{figure*}[htbp]
    \centering
    \includegraphics[width=.9\linewidth]{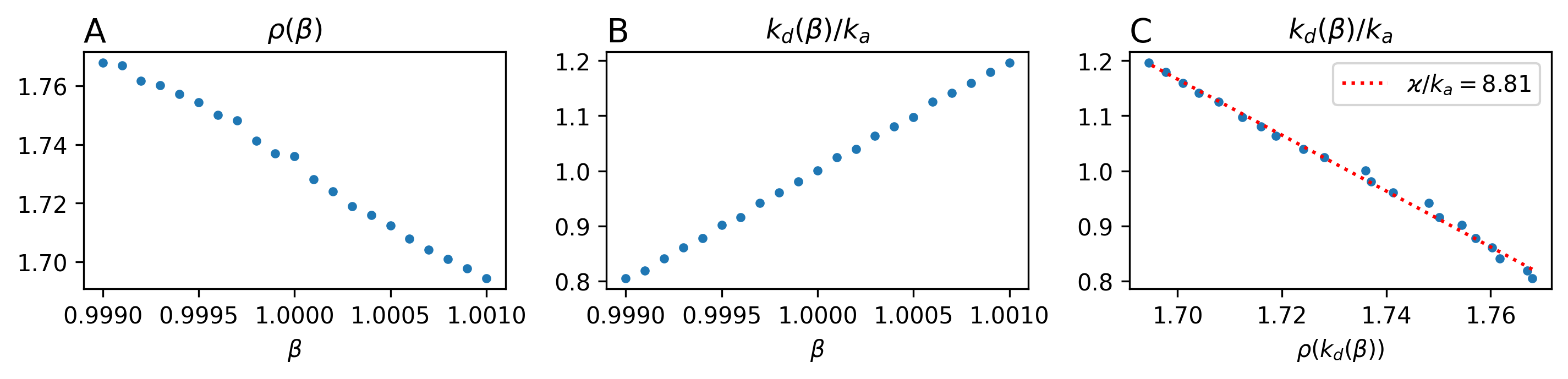}
    \caption[Perturbation-response in agent-based simulations.]{
        \textbf{Perturbation-response in agent-based simulations.}
        Example how response is measured via perturbation, where $\beta$ is the rescaling coefficient of volume change and 
        (A) density and
        (B) division rate variation are shown as a function of rescaling.
        (C) Fitting the response rate $\varkappa_\beta$ in Eq.\,\ref{eq:ResponseDensity} to data.
    }
    \label{fig:app:Fig4_2PGsimu_compressionexpansionresponse}
\end{figure*}
The two-particle growth (2PG) model of Refs.~\cite{Basan2011dpd,Podewitz2015} and has been adapted in Refs.~\cite{Ganai2019,Buescher2020a,Buescher2020b,Kraemer2024} to a variety of different models.

Each cell is described by two particles, which repel each other via an active growth force
\begin{equation}
	F_{ij}^{G} = \frac{G}{\left( r_{ij} + r_0 \right)^2} \hat{\boldsymbol{r}}_{ij}
\end{equation}
with unit vector $\hat{\boldsymbol{r}}_{ij}$, distance $r_{ij}$ between the two particles, a constant $r_0$, and growth strength $G$. 

To prevent overlap of cells, particles of different cells interact via a soft repulsive volume-exclusion force
\begin{equation}
	\boldsymbol{F}_{ik}^{V} 
	= f_0 \left( \frac{R_{PP}^5}{r_{ik}^5} - 1 \right) \hat{\boldsymbol{r}}_{ik}
	\text{, for }r_{ik} < R_{PP}
\end{equation}
with exclusion strength $f_0$ and interaction length $R_{PP}$, which sets the length scale of our simulations.
Further, cells in contact interact via an attractive adhesion force of the form 
\begin{equation}
	\boldsymbol{F}_{ik}^{A} = -f_1 \hat{\boldsymbol{r}}_{ik}
	\text{, for }r_{ik} < R_{PP}
\end{equation}
with adhesion strength $f_1$.
We employ a dissipative particle dynamics-type thermostat~\cite{hoogerbrugge_simulating_1992}, with an effective temperature $T$, to account for energy dissipation 
\begin{equation}
	F^{D}_{ij} = \gamma \cdot w^D(r_{ij}) \cdot \left(\boldsymbol{v}_{ij}\cdot \hat{\boldsymbol{r}}_{ij}\right) \hat{\boldsymbol{r}}_{ij}
\end{equation}
and random fluctuations
\begin{equation}
	F^{R}_{ij} = \sigma w^R(r_{ij}) \xi \cdot \hat{\boldsymbol{r}}_{ij}
\end{equation}
where $\gamma$ and $\sigma$ and the weight functions $w^{D/R}(r_{ij})$ are related to fulfill the fluctuation-dissipation theorem~\cite{espanol_statistical_1995}.
Also, background dissipation is taken into account as
\begin{equation}
	\boldsymbol{F}^B_i = -\gamma_b\boldsymbol{v}_i.
\end{equation}

A self-consistent velocity-Verlet algorithm~\cite{pagonabarraga_self-consistent_1998} is implemented to integrate the equations of motion, and all simulations were performed with periodic boundary conditions.

Cell division is performed, when the two particles of one cell are separated by a critical threshold $R_{ct}$. A new particle is randomly placed near each original particle in a distance $R_{d}$, and the two new particle pairs form the two daughter cells.
Cells are stochastically removed from the simulation with a finite apoptosis rate $k_a$. Time is measured in terms of the inverse apoptosis rate, and referred to as "generation".
The standard parameter set for our simulations are given in Tab.~\ref{tab:standard_parameter_set}. 
However, not all of these simulation parameters, have a direct conversion to physical units. As discussed in \cite{Basan2011dpd} one has to chose well defined measurable quantities, such as apoptosis rate and range of the pair potential as rescaling units for inverse time and particle diameter, to allow for conversion to physical units and comparison with experiments.
Simulations are performed in a two-dimensional squared area with periodic boundary conditions.
\begin{table}[ht]
    \centering
    \caption{2PG tissue simulation parameters (standard parameter set).}
    \begin{tabular}{lll}
    \toprule
        Parameter & Symbol & Value \\
    \midrule
        Time step & $\Delta t$ & 0.001 \\
        Pair potential interaction range & $R_{PP}$ & 1 \\
        Cell expansion constant & $r_0$ & 1 \\
        Division distance threshold & $R_{ct}$ & 0.8 \\
        New cell particle initial distance & $R_{d}$ & $10^{-5}$ \\
        Growth-force strength & $G$ & 28 \\
        Mass & $m$ & 1 \\
        Intra-cell dissipation coefficient & $\gamma_c$ & 100 \\
        Inter-cell dissipation coefficient & $\gamma_t$ & 50 \\
        Background dissipation coefficient & $\gamma_{bg}$ & 0.1 \\
        Apoptosis rate & $k_a$ & 0.01 \\
        Noise intensity & $k_BT$ & 0.1 \\
        Repulsive cell-cell potential coefficient & $f_0$ & 2.39566 \\
        Attractive cell-cell potential coefficient & $f_1$ & 7.0 \\
    \bottomrule
    \end{tabular}
    \label{tab:standard_parameter_set}
\end{table}

\paragraph{Compression expansion module} \label{sec:SI:SimulationModels:2PG:compexp}
\begin{figure*}[htbp] 
    \centering
    \includegraphics[width=.9\linewidth]{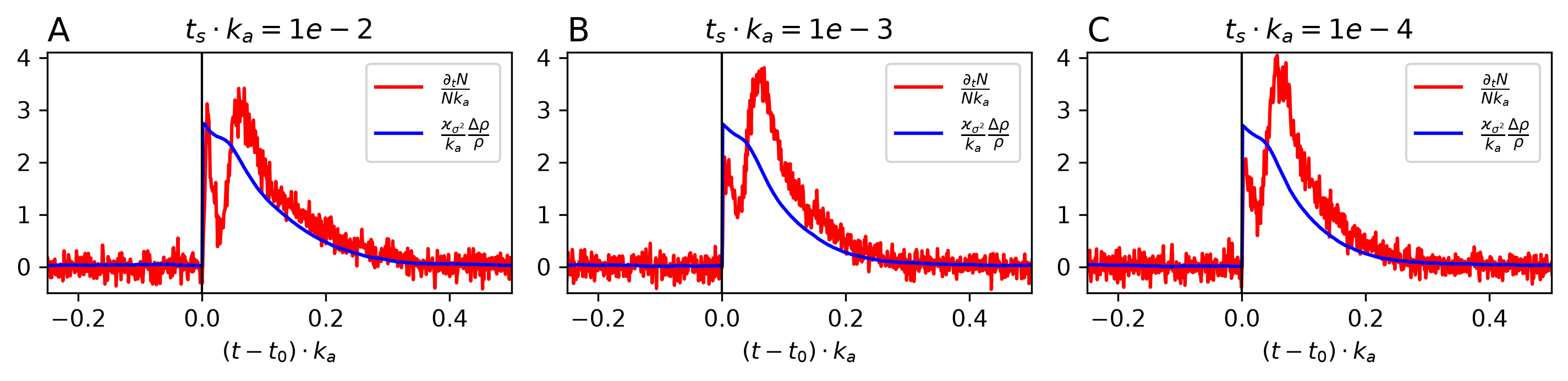}
    \caption[Stretch response rate dependence.]{
        \textbf{Stretch response rate dependence.}
        2PG simulations for three different stretch duration.
        In all simulations, 20\% stretch is realized, however, over different time intervals.
    }
    \label{Fig:SI:stretchresponse}
\end{figure*}
In mechanical tissue simulations, the response coefficient is a priori not known, but results from the mechanical properties of cells and their local interactions.
For monoclonal tissues, local configuration can be expected to play a minor role, but for more complex multicellular tissues this may become relevant.
We compare the result for measurements of $\varkappa_{\sigma^2}$, obtained from the fluctuation response methods, and $\varkappa_{\beta}$, the response to volume rescaling.

The volume rescaling method is motivated by common experimental techniques. 
For example cell spheroids can be grown under increased pressure, by placing them in a semipermeable membrane and controlling the osmotic pressure by the dextrane concentration of the surrounding medium~\cite{Montel2011}.
Also, cells grown on an elastic substrate can be tested for response to stretch and release of the underlying substrate.
We transfer these methods to simulations, where the simulation volume is rescaled at a constant rate $\beta$.
Cell center of mass positions are updated by the same factor, while cell sizes remain unchanged. The position of a cell is given by
\begin{equation}
    \boldsymbol{r}_{new} = \beta \cdot \boldsymbol{r}_{old}^{PBC},
\end{equation}
were $\boldsymbol{r}_{old}^{PBC}$ is the position of the particle in the unit box of our simulations, ranging from $(0,L_{xy}(t))$.

This allows to observe the response of the division rate as a consequence of reduced~($\beta>1$) or increased~($\beta<1$) density (see Fig.\,\ref{fig:app:Fig4_2PGsimu_compressionexpansionresponse}).
While in principle this method can be applied for infinite time, large volumes come with increased computational time, which limits our experiments. Thus, data is averaged over independent rescaling runs to improve statistics.
Plotting the division rate as a function of density, we find the expected linear relation, where the slope equals the response coefficient $\varkappa_{\beta}$ (see Fig.\,\ref{fig:app:Fig4_2PGsimu_compressionexpansionresponse}).

Although the exact value of $\varkappa$ is not known, certain model parameters allow for a physical interpretation, which can be confirmed qualitatively.
The active growth force of cells is controlled by the growth coefficient $G$. With an increase of $G$ one can associate a faster response to cell loss. Thus, higher value of $\varkappa$.
This is confirmed by our measurements.
For too small values of $G$, the tissue loses confluency, which in particular affects the rescaling measurement, which requires a confluent tissue even under expansion.
With increasing growth coefficient, both methods reach a plateau, since fluidization gets lost and a crystal-like phase is realized. This already indicates the limit of biological relevant parameters.

Since the beginning of computational modeling of cellular systems, cell adhesion is known to be a crucial factor, which drives cell sorting or tissue fluidization~\cite{Graner1992}.
We find, that higher adhesion reduces the response coefficient.
Tightly coupled cells can not as easily rearrange to respond to cell loss, thus response is smaller for high adhesion.
Lower adhesion supports fluidization of cells, and an increase of the response coefficient.

In the mechanical model, time is measured in generations, set by the fixed apoptosis rate $k_a$. 
On the other hand, apoptosis is the relevant process, which initiates fluidization of tissues~\cite{Basan2011dpd}.
As discussed above, fluidization is relevant for quick response, and we can confirm that increasing $k_a$ comes with an increase of $\varkappa$. 
Note, that the ratio $\varkappa/k_a$ stays approximately constant.

The same rescaling method is applied to the stretch-experiments, however, only for a short time with much larger scaling factor.

\subsubsection{Voronoi membrane potential growth (VMPG) model}\label{sec:SI:SimulationModels:VMPG}
As a more sophisticated agent-based model, we adapt the simulations originally published by Mukherjee~et~al.~\cite{Mukherjee2026}, where a more detailed description is given.
Again, simulations are performed in 2D with periodic boundary conditions.
When solving the mechano-electro-osmotic model, a constant cell height is assumed, as cell height fluctuations are only expected to be small.
Cells are represented by their center position, and cell shape is constructed via Voronoi triangulation.
The mechano-electro-osmotic model is solved for each cell to calculate the ion-concentrations, membrane potential $V$ and pressure $P$.
These are then used to calculate forces on a cell from the gradient of an energy functional, given by
\begin{equation}
    E_{cell} = P \cdot V + T \cdot (p - p_0)^2,
\end{equation}
with tension coefficient $T$ and current perimeter $p$ and target perimeter $p_0$.
In addition, frictional forces with the substrate $\boldsymbol{f}_{sub} = -\gamma \boldsymbol{v}$ act on cells moving with velocity $\textbf{v}$ and friction coefficient $\gamma$.
Cellular biomass degrades with a constant rate and is produced proportional to the membrane potential $U$, until a threshold value for the potential is reached.
Cell division is implemented as an \textit{adder}-model: A cell divides, if a constant increase in biomass is reached. This value is decided at the time of division and drawn from a Gaussian distribution.
Cell death is a stochastic process with half-life time of five days, except for high potentials succeeding a value of $-70mV$, when the rate is doubled resulting in a half-life time of 2.5~days.
All simulations parameters are chosen the same as in \cite{Mukherjee2026}, where they were calibrated against experimental data.

Rescaling for stretch-experiments is performed in the same fashion as for the 2PG simulations.

\subsubsection{Stretch Response}\label{sec:SI:stretchresponse}

In 2PG stretch simulations, we find a small increase in the current division rate directly after applying stretch.
We speculate that this peak is a result from the stretch-method implemented in simulations.
Stretch is not instantaneous, but carried out over a short time.
We test how the peak changes for different stretch times, ranging over two orders of magnitude from $t\cdot k_a = 1e-4$ (10 integration steps) to $t\cdot k_a = 1e-2$ (1000 integration steps) we perform stretch experiments, all resulting in 20\% stretch (see Fig.\,\ref{Fig:SI:stretchresponse}).
We find, that for the slowest process, the peak is most pronounced, while no additional saturation is found for even shorter times.
Note that only cell position and not cell size is rescaled during the stretch.
A possible explanation is, that adhesive forces cause additional growth, which is more pronounced if cells pull on each other continuously.
While here this has to be understood as an artifact of the simulation setup, real cells are expected to show physiological change when stress is applied externally.
Thus, highlighting the need of non-invasive methods.
In the main text figure~\ref{fig:Fig3_2PGsimu_fluctuationresponse}, data for the intermediate stretch time is shown.

\subsubsection{Mean back relaxation (MBR)}\label{sec:SI:MBR}
\begin{figure*}[htbp]
    \centering
    \includegraphics[width=.7\linewidth]{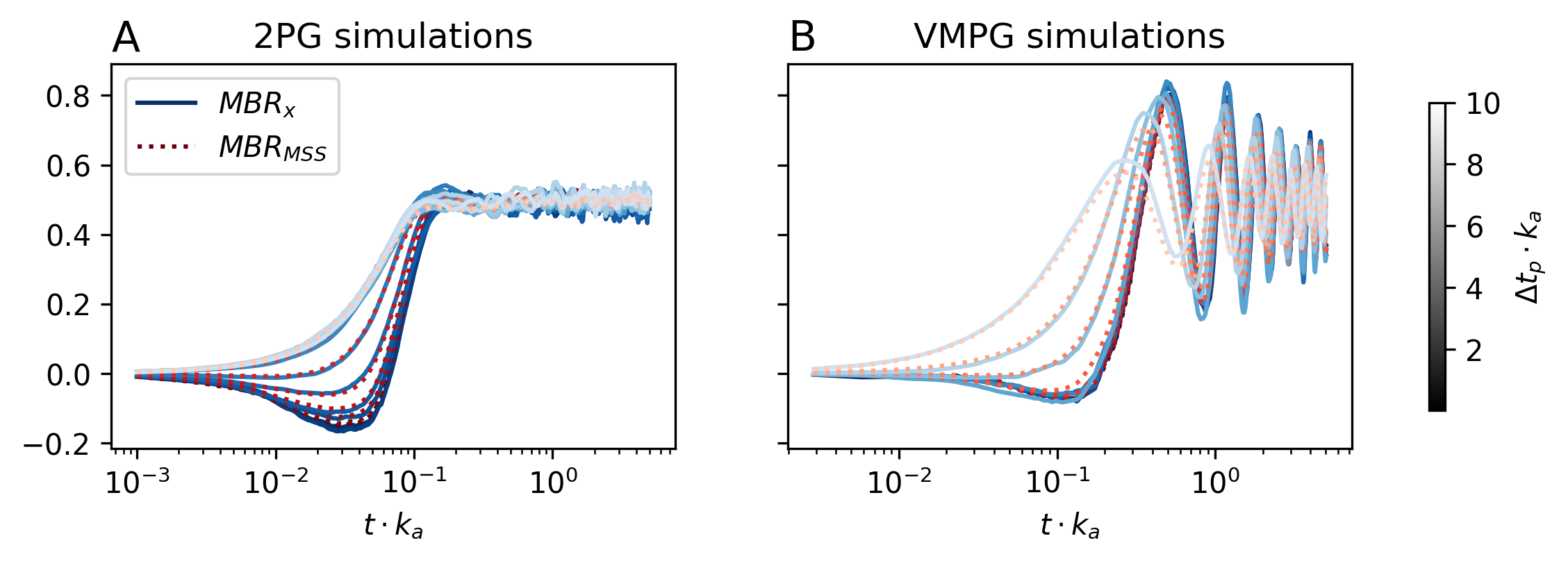}
    \caption[Mean back relaxation.]{
    \textbf{Mean back relaxation.}
        Mean back relaxation function ($MBR$) over forward time $\Delta t_f$ for color-coded backward times $\Delta t_p$ (dark: short displacement; bright: large displacement).
        Solid line is calculated directly from data by Eq.\,\ref{eq:SI:MBR} with $l_c = 10^{-5}$.
        and the dotted line via the $MSS$ following Eq.\,\ref{eq:SI:MBRMSS}.
        (A) shows data for 2PG simulations and 
        (B) data for the VMPG simulations.
    }
    \label{fig:SI:MBR}
\end{figure*}

All previously discussed dynamic methods correlate the tissue population at two time points. 
Recently a related quantity, the mean back relaxation ($MBR$), has been introduced, which correlates three time points of a trajectory~\cite{Knotz2024}.
It is defined by the average future displacement divided by the past displacement and given by
\begin{equation}
\begin{aligned}
    &MBR_x(\Delta t_f, \Delta t_p, l_c) =
    \\
    &\quad\left\langle 
        \frac{
            x(t+\Delta t_f) - x(t)
        }{
            x(t) - x(t-\Delta t_p)
        }
        \theta_{l_c}(\left| x(t) - x(t-\Delta t_p) \right|)
    \right\rangle,
\end{aligned}
    \label{eq:SI:MBR}
\end{equation}
where $\Delta t_f$ and $\Delta t_p$ are the time displacements to the future and past, respectively. 
$l_c$ is a cutoff value, which prevents division by zero, and $\theta_{l_c}$ a Heaviside function.
An alternative representative of the $MBR$ can be given for Gaussian systems in the form
\begin{equation}
\begin{aligned}
    &MBR_{MSS}(\Delta t_f, \Delta t_p) = 
    \\
    &\quad\frac{1}{2} \left(
    1 - \frac{
            MSS(\Delta t_f + \Delta t_p) - MSS(\Delta t_f)
        }{
            MSS(\Delta t_p)
        }
    \right),
\end{aligned}
    \label{eq:SI:MBRMSS}
\end{equation}
which is independent of the cutoff $l_c$ and relates the $MBR$ to the $MSS$.
In general, this three-point correlation function provides a condensed insight in the dynamic of the system and has been shown to be in particular helpful to probe the non-equilibrium nature of a system with passive measurements~\cite{Muenker2024}.
Both agent-based models studied in this work do result in a Gaussian cell density distribution, thus, both methods to calculate the $MBR$ are equivalent (see Fig.\,\ref{fig:SI:MBR}).
Further, we do not find strong indication for broken detailed balance in the 2PG simulations, as the $MBR$ converges to 1/2, while for the VMPG simulations, oscillations make it difficult to correctly identify the long-term limit.

\bibliography{sample}

@article{BergSoerensen2004,
  title = {Power spectrum analysis for optical tweezers},
  volume = {75},
  ISSN = {1089-7623},
  url = {http://dx.doi.org/10.1063/1.1645654},
  DOI = {10.1063/1.1645654},
  number = {3},
  journal = {Review of Scientific Instruments},
  publisher = {AIP Publishing},
  author = {Berg-Sørensen,  Kirstine and Flyvbjerg,  Henrik},
  year = {2004},
  month = mar,
  pages = {594–612}
}

@article{Meyer2023,
  title = {Stochastic processes in a confining harmonic potential in the presence of static and dynamic measurement noise},
  volume = {25},
  ISSN = {1367-2630},
  url = {http://dx.doi.org/10.1088/1367-2630/acd94f},
  DOI = {10.1088/1367-2630/acd94f},
  number = {6},
  journal = {New Journal of Physics},
  publisher = {IOP Publishing},
  author = {Meyer,  Philipp G and Metzler,  Ralf},
  year = {2023},
  month = jun,
  pages = {063003}
}

@article{Meyer2021,
  title = {Time reversal symmetry and the difference between relaxations and building-up periods},
  volume = {104},
  ISSN = {2470-0053},
  url = {http://dx.doi.org/10.1103/PhysRevE.104.024208},
  DOI = {10.1103/physreve.104.024208},
  number = {2},
  journal = {Physical Review E},
  publisher = {American Physical Society (APS)},
  author = {Meyer,  Philipp G. and Kantz,  Holger},
  year = {2021},
  month = aug 
}

@article{Cherstvy2018,
  title = {Time averages and their statistical variation for the Ornstein-Uhlenbeck process: Role of initial particle distributions and relaxation to stationarity},
  volume = {98},
  ISSN = {2470-0053},
  url = {http://dx.doi.org/10.1103/PhysRevE.98.022134},
  DOI = {10.1103/physreve.98.022134},
  number = {2},
  journal = {Physical Review E},
  publisher = {American Physical Society (APS)},
  author = {Cherstvy,  Andrey G. and Thapa,  Samudrajit and Mardoukhi,  Yousof and Chechkin,  Aleksei V. and Metzler,  Ralf},
  year = {2018},
  month = aug 
}

@article{Mukherjee2026,
  title = {Membrane potential mediates the cellular response to mechanical pressure},
  volume = {189},
  ISSN = {0092-8674},
  url = {http://dx.doi.org/10.1016/j.cell.2025.11.004},
  DOI = {10.1016/j.cell.2025.11.004},
  number = {1},
  journal = {Cell},
  publisher = {Elsevier BV},
  author = {Mukherjee,  Avik and Huang,  Yanqing and Elgeti,  Jens and Oh,  Seungeun and Abreu,  Jose G. and Ammar,  Leander and Neliat,  Anjali R. and Sch\"{u}ttler,  Janik and Su,  Dan-Dan and Dupre,  Christophe and Benites,  Nina Catherine and Liu,  Xili and Peshkin,  Leonid and Barboiu,  Mihail and Stocker,  Hugo and Kirschner,  Marc W. and Basan,  Markus},
  year = {2026},
  month = jan,
  pages = {143--160.e22}
}

@article{Hallatschek2023,
  title = {Proliferating active matter},
  volume = {5},
  ISSN = {2522-5820},
  url = {http://dx.doi.org/10.1038/s42254-023-00593-0},
  DOI = {10.1038/s42254-023-00593-0},
  number = {7},
  journal = {Nature Reviews Physics},
  publisher = {Springer Science and Business Media LLC},
  author = {Hallatschek,  Oskar and Datta,  Sujit S. and Drescher,  Knut and Dunkel,  J\"{o}rn and Elgeti,  Jens and Waclaw,  Bartek and Wingreen,  Ned S.},
  year = {2023},
  month = may,
  pages = {407–419}
}

@article{Basan2009,
  title = {Homeostatic competition drives tumor growth and metastasis nucleation},
  volume = {3},
  ISSN = {1955-2068},
  url = {http://dx.doi.org/10.2976/1.3086732},
  DOI = {10.2976/1.3086732},
  number = {4},
  journal = {HFSP Journal},
  publisher = {Informa UK Limited},
  author = {Basan,  Markus and Risler,  Thomas and Joanny,  Jean‐Fran\c{c}ois and Sastre‐Garau,  Xavier and Prost,  Jacques},
  year = {2009},
  month = aug,
  pages = {265–272}
}

@article{Montel2011,
  title = {Stress Clamp Experiments on Multicellular Tumor Spheroids},
  volume = {107},
  ISSN = {1079-7114},
  url = {http://dx.doi.org/10.1103/PhysRevLett.107.188102},
  DOI = {10.1103/physrevlett.107.188102},
  number = {18},
  journal = {Physical Review Letters},
  publisher = {American Physical Society (APS)},
  author = {Montel,  Fabien and Delarue,  Morgan and Elgeti,  Jens and Malaquin,  Laurent and Basan,  Markus and Risler,  Thomas and Cabane,  Bernard and Vignjevic,  Danijela and Prost,  Jacques and Cappello,  Giovanni and Joanny,  Jean-Fran\c{c}ois},
  year = {2011},
  month = oct 
}

@article{Jeon2012,
  title = {Inequivalence of time and ensemble averages in ergodic systems: Exponential versus power-law relaxation in confinement},
  volume = {85},
  ISSN = {1550-2376},
  url = {http://dx.doi.org/10.1103/PhysRevE.85.021147},
  DOI = {10.1103/physreve.85.021147},
  number = {2},
  journal = {Physical Review E},
  publisher = {American Physical Society (APS)},
  author = {Jeon,  Jae-Hyung and Metzler,  Ralf},
  year = {2012},
  month = feb 
}

@article{Graner1992,
  title = {Simulation of biological cell sorting using a two-dimensional extended Potts model},
  volume = {69},
  ISSN = {0031-9007},
  url = {http://dx.doi.org/10.1103/PhysRevLett.69.2013},
  DOI = {10.1103/physrevlett.69.2013},
  number = {13},
  journal = {Physical Review Letters},
  publisher = {American Physical Society (APS)},
  author = {Graner,  Fran\c{c}ois and Glazier,  James A.},
  year = {1992},
  month = sep,
  pages = {2013–2016}
}

@article{Ai2003,
  title = {Correlated noise in a logistic growth model},
  volume = {67},
  ISSN = {1095-3787},
  url = {http://dx.doi.org/10.1103/PhysRevE.67.022903},
  DOI = {10.1103/physreve.67.022903},
  number = {2},
  journal = {Physical Review E},
  publisher = {American Physical Society (APS)},
  author = {Ai,  Bao-Quan and Wang,  Xian-Ju and Liu,  Guo-Tao and Liu,  Liang-Gang},
  year = {2003},
  month = feb 
}

@article{Ranft2010,
  title = {Fluidization of tissues by cell division and apoptosis},
  volume = {107},
  ISSN = {1091-6490},
  url = {http://dx.doi.org/10.1073/pnas.1011086107},
  DOI = {10.1073/pnas.1011086107},
  number = {49},
  journal = {Proceedings of the National Academy of Sciences},
  publisher = {Proceedings of the National Academy of Sciences},
  author = {Ranft,  Jonas and Basan,  Markus and Elgeti,  Jens and Joanny,  Jean-Fran\c{c}ois and Prost,  Jacques and J\"{u}licher,  Frank},
  year = {2010},
  month = nov,
  pages = {20863–20868}
}

@article{Basan2011dpd,
  title = {Dissipative particle dynamics simulations for biological tissues: rheology and competition},
  volume = {8},
  ISSN = {1478-3975},
  url = {http://dx.doi.org/10.1088/1478-3975/8/2/026014},
  DOI = {10.1088/1478-3975/8/2/026014},
  number = {2},
  journal = {Physical Biology},
  publisher = {IOP Publishing},
  author = {Basan,  Markus and Prost,  Jacques and Joanny,  Jean-Fran\c{c}ois and Elgeti,  Jens},
  year = {2011},
  month = apr,
  pages = {026014}
}

@article{Podewitz2015,
  title = {Tissue homeostasis: A tensile state},
  volume = {109},
  ISSN = {1286-4854},
  url = {http://dx.doi.org/10.1209/0295-5075/109/58005},
  DOI = {10.1209/0295-5075/109/58005},
  number = {5},
  journal = {EPL (Europhysics Letters)},
  publisher = {IOP Publishing},
  author = {Podewitz,  N. and Delarue,  M. and Elgeti,  J.},
  year = {2015},
  month = mar,
  pages = {58005}
}

@article{Ganai2019,
  title = {Mechanics of tissue competition: interfaces stabilize coexistence},
  volume = {21},
  ISSN = {1367-2630},
  url = {http://dx.doi.org/10.1088/1367-2630/ab2475},
  DOI = {10.1088/1367-2630/ab2475},
  number = {6},
  journal = {New Journal of Physics},
  publisher = {IOP Publishing},
  author = {Ganai,  Nirmalendu and B\"{u}scher,  Tobias and Gompper,  Gerhard and Elgeti,  Jens},
  year = {2019},
  month = jun,
  pages = {063017}
}

@article{Buescher2020a,
  title = {Tissue evolution: mechanical interplay of adhesion,  pressure,  and heterogeneity},
  volume = {22},
  ISSN = {1367-2630},
  url = {http://dx.doi.org/10.1088/1367-2630/ab74a5},
  DOI = {10.1088/1367-2630/ab74a5},
  number = {3},
  journal = {New Journal of Physics},
  publisher = {IOP Publishing},
  author = {B\"{u}scher,  Tobias and Ganai,  Nirmalendu and Gompper,  Gerhard and Elgeti,  Jens},
  year = {2020},
  month = mar,
  pages = {033048}
}

@article{Buescher2020b,
  title = {Instability and fingering of interfaces in growing tissue},
  volume = {22},
  ISSN = {1367-2630},
  url = {http://dx.doi.org/10.1088/1367-2630/ab9e88},
  DOI = {10.1088/1367-2630/ab9e88},
  number = {8},
  journal = {New Journal of Physics},
  publisher = {IOP Publishing},
  author = {B\"{u}scher,  Tobias and Diez,  Angel L and Gompper,  Gerhard and Elgeti,  Jens},
  year = {2020},
  month = aug,
  pages = {083005}
}

@article{Kraemer2024,
  title = {Mechanically-driven stem cell separation in tissues caused by proliferating daughter cells},
  volume = {16},
  ISSN = {2542-4653},
  url = {http://dx.doi.org/10.21468/SciPostPhys.16.4.097},
  DOI = {10.21468/scipostphys.16.4.097},
  number = {4},
  journal = {SciPost Physics},
  publisher = {Stichting SciPost},
  author = {Kr\"{a}mer,  Johannes C. and Hannezo,  Edouard and Gompper,  Gerhard and Elgeti,  Jens},
  year = {2024},
  month = apr 
}

@article{hoogerbrugge_simulating_1992,
	title = {Simulating Microscopic Hydrodynamic Phenomena with Dissipative Particle Dynamics},
	volume = {19},
	issn = {0295-5075, 1286-4854},
	url = {https://iopscience.iop.org/article/10.1209/0295-5075/19/3/001},
	doi = {10.1209/0295-5075/19/3/001},
	number = {3},
	urldate = {2023-07-18},
	journal = {Europhysics Letters (EPL)},
	author = {Hoogerbrugge, P. J and Koelman, J. M. V. A},
	month = jun,
	year = {1992},
	pages = {155--160},
}

@article{pagonabarraga_self-consistent_1998,
	title = {Self-consistent dissipative particle dynamics algorithm},
	volume = {42},
	issn = {0295-5075, 1286-4854},
	url = {https://iopscience.iop.org/article/10.1209/epl/i1998-00258-6},
	doi = {10.1209/epl/i1998-00258-6},
	number = {4},
	urldate = {2023-07-18},
	journal = {Europhysics Letters (EPL)},
	author = {Pagonabarraga, I and Hagen, M. H. J and Frenkel, D},
	month = may,
	year = {1998},
	pages = {377--382},
}

@article{espanol_statistical_1995,
	title = {Statistical Mechanics of Dissipative Particle Dynamics},
	volume = {30},
	issn = {0295-5075, 1286-4854},
	url = {https://iopscience.iop.org/article/10.1209/0295-5075/30/4/001},
	doi = {10.1209/0295-5075/30/4/001},
	number = {4},
	urldate = {2023-07-18},
	journal = {Europhysics Letters (EPL)},
	author = {Español, P and Warren, P},
	month = may,
	year = {1995},
	pages = {191--196},
}

@article{Knotz2024,
  title = {Mean back relaxation for position and densities},
  volume = {110},
  ISSN = {2470-0053},
  url = {http://dx.doi.org/10.1103/PhysRevE.110.044137},
  DOI = {10.1103/physreve.110.044137},
  number = {4},
  journal = {Physical Review E},
  publisher = {American Physical Society (APS)},
  author = {Knotz,  Gabriel and Kr\"{u}ger,  Matthias},
  year = {2024},
  month = oct,
}

@article{Muenker2024,
  title = {Accessing activity and viscoelastic properties of artificial and living systems from passive measurement},
  volume = {23},
  ISSN = {1476-4660},
  url = {http://dx.doi.org/10.1038/s41563-024-01957-2},
  DOI = {10.1038/s41563-024-01957-2},
  number = {9},
  journal = {Nature Materials},
  publisher = {Springer Science and Business Media LLC},
  author = {Muenker,  Till M. and Knotz,  Gabriel and Kr\"{u}ger,  Matthias and Betz,  Timo},
  year = {2024},
  month = jul,
  pages = {1283–1291},
}

@article{Turlier2016,
  title = {Equilibrium physics breakdown reveals the active nature of red blood cell flickering},
  volume = {12},
  ISSN = {1745-2481},
  url = {http://dx.doi.org/10.1038/nphys3621},
  DOI = {10.1038/nphys3621},
  number = {5},
  journal = {Nature Physics},
  publisher = {Springer Science and Business Media LLC},
  author = {Turlier,  H. and Fedosov,  D. A. and Audoly,  B. and Auth,  T. and Gov,  N. S. and Sykes,  C. and Joanny,  J.-F. and Gompper,  G. and Betz,  T.},
  year = {2016},
  month = jan,
  pages = {513–519}
}

@article{Shraiman2005,
  title = {Mechanical feedback as a possible regulator of tissue growth},
  volume = {102},
  ISSN = {1091-6490},
  url = {http://dx.doi.org/10.1073/pnas.0404782102},
  DOI = {10.1073/pnas.0404782102},
  number = {9},
  journal = {Proceedings of the National Academy of Sciences},
  publisher = {Proceedings of the National Academy of Sciences},
  author = {Shraiman,  Boris I.},
  year = {2005},
  month = feb,
  pages = {3318–3323}
}

@article{Hannezo2019,
  title = {Mechanochemical Feedback Loops in Development and Disease},
  volume = {178},
  ISSN = {0092-8674},
  url = {http://dx.doi.org/10.1016/j.cell.2019.05.052},
  DOI = {10.1016/j.cell.2019.05.052},
  number = {1},
  journal = {Cell},
  publisher = {Elsevier BV},
  author = {Hannezo,  Edouard and Heisenberg,  Carl-Philipp},
  year = {2019},
  month = Jun,
  pages = {12–25}
}

@article{Alasaadi2024R,
  title = {Mechanically guided cell fate determination in early development},
  volume = {81},
  ISSN = {1420-9071},
  url = {http://dx.doi.org/10.1007/s00018-024-05272-6},
  DOI = {10.1007/s00018-024-05272-6},
  number = {1},
  journal = {Cellular and Molecular Life Sciences},
  publisher = {Springer Science and Business Media LLC},
  author = {Alasaadi,  Delan N. and Mayor,  Roberto},
  year = {2024},
  month = may 
}

@article{Ferrai2024,
  title = {Mechanotransduction in stem cells},
  volume = {103},
  ISSN = {0171-9335},
  url = {http://dx.doi.org/10.1016/j.ejcb.2024.151417},
  DOI = {10.1016/j.ejcb.2024.151417},
  number = {2},
  journal = {European Journal of Cell Biology},
  publisher = {Elsevier BV},
  author = {Ferrai,  Carmelo and Schulte,  Carsten},
  year = {2024},
  month = jun,
  pages = {151417}
}

@article{Callen1951,
  title = {Irreversibility and Generalized Noise},
  volume = {83},
  ISSN = {0031-899X},
  url = {http://dx.doi.org/10.1103/PhysRev.83.34},
  DOI = {10.1103/physrev.83.34},
  number = {1},
  journal = {Physical Review},
  publisher = {American Physical Society (APS)},
  author = {Callen,  Herbert B. and Welton,  Theodore A.},
  year = {1951},
  month = Jul,
  pages = {34–40}
}

@article{Kirkpatrick2025,
  title = {Fluctuation-response relation in nonequilibrium systems and active matter},
  volume = {111},
  ISSN = {2470-0053},
  url = {http://dx.doi.org/10.1103/PhysRevE.111.014102},
  DOI = {10.1103/physreve.111.014102},
  number = {1},
  journal = {Physical Review E},
  publisher = {American Physical Society (APS)},
  author = {Kirkpatrick,  T. R. and Belitz,  D.},
  year = {2025},
  month = Jan 
}

@article{Kubo1966,
  title = {The fluctuation-dissipation theorem},
  volume = {29},
  ISSN = {0034-4885},
  url = {http://dx.doi.org/10.1088/0034-4885/29/1/306},
  DOI = {10.1088/0034-4885/29/1/306},
  number = {1},
  journal = {Reports on Progress in Physics},
  publisher = {IOP Publishing},
  author = {Kubo,  R},
  year = {1966},
  month = Jan,
  pages = {255–284}
}

@article{Sato2003,
  title = {On the relation between fluctuation and response in biological systems},
  volume = {100},
  ISSN = {1091-6490},
  url = {http://dx.doi.org/10.1073/pnas.2334996100},
  DOI = {10.1073/pnas.2334996100},
  number = {24},
  journal = {Proceedings of the National Academy of Sciences},
  publisher = {Proceedings of the National Academy of Sciences},
  author = {Sato,  Katsuhiko and Ito,  Yoichiro and Yomo,  Tetsuya and Kaneko,  Kunihiko},
  year = {2003},
  month = Nov,
  pages = {14086–14090}
}

@article{Constantinou2023,
  title = {Cell-stretching devices: advances and challenges in biomedical research and live-cell imaging},
  volume = {41},
  ISSN = {0167-7799},
  url = {http://dx.doi.org/10.1016/j.tibtech.2022.12.009},
  DOI = {10.1016/j.tibtech.2022.12.009},
  number = {7},
  journal = {Trends in Biotechnology},
  publisher = {Elsevier BV},
  author = {Constantinou,  Iordania and Bastounis,  Effie E.},
  year = {2023},
  month = Jul,
  pages = {939–950}
}

@article{Bighi2024,
  title = {Cell stretching devices integrated with live cell imaging: a powerful approach to study how cells react to mechanical cues},
  volume = {7},
  ISSN = {2516-1091},
  url = {http://dx.doi.org/10.1088/2516-1091/ad9699},
  DOI = {10.1088/2516-1091/ad9699},
  number = {1},
  journal = {Progress in Biomedical Engineering},
  publisher = {IOP Publishing},
  author = {Bighi,  Beatrice and Ragazzini,  Gregorio and Gallerani,  Alessia and Mescola,  Andrea and Scagliarini,  Chiara and Zannini,  Chiara and Marcuzzi,  Martina and Olivi,  Elena and Cavallini,  Claudia and Tassinari,  Riccardo and Bianchi,  Michele and Corsi,  Lorenzo and Ventura,  Carlo and Alessandrini,  Andrea},
  year = {2024},
  month = Dec,
  pages = {012005}
}

@article{Aragona2020,
  title = {Mechanisms of stretch-mediated skin expansion at single-cell resolution},
  volume = {584},
  ISSN = {1476-4687},
  url = {http://dx.doi.org/10.1038/s41586-020-2555-7},
  DOI = {10.1038/s41586-020-2555-7},
  number = {7820},
  journal = {Nature},
  publisher = {Springer Science and Business Media LLC},
  author = {Aragona,  Mariaceleste and Sifrim,  Alejandro and Malfait,  Milan and Song,  Yura and Van Herck,  Jens and Dekoninck,  Sophie and Gargouri,  Souhir and Lapouge,  Gaëlle and Swedlund,  Benjamin and Dubois,  Christine and Baatsen,  Pieter and Vints,  Katlijn and Han,  Seungmin and Tissir,  Fadel and Voet,  Thierry and Simons,  Benjamin D. and Blanpain,  Cédric},
  year = {2020},
  month = Jul,
  pages = {268–273}
}

@article{Marchetti2013,
  title = {Hydrodynamics of soft active matter},
  volume = {85},
  ISSN = {1539-0756},
  url = {http://dx.doi.org/10.1103/RevModPhys.85.1143},
  DOI = {10.1103/revmodphys.85.1143},
  number = {3},
  journal = {Reviews of Modern Physics},
  publisher = {American Physical Society (APS)},
  author = {Marchetti,  M. C. and Joanny,  J. F. and Ramaswamy,  S. and Liverpool,  T. B. and Prost,  J. and Rao,  Madan and Simha,  R. Aditi},
  year = {2013},
  month = jul,
  pages = {1143–1189}
}

@article{Ramaswamy2010,
  title = {The Mechanics and Statistics of Active Matter},
  volume = {1},
  ISSN = {1947-5462},
  url = {http://dx.doi.org/10.1146/annurev-conmatphys-070909-104101},
  DOI = {10.1146/annurev-conmatphys-070909-104101},
  number = {1},
  journal = {Annual Review of Condensed Matter Physics},
  publisher = {Annual Reviews},
  author = {Ramaswamy,  Sriram},
  year = {2010},
  month = aug,
  pages = {323–345}
}

@inbook{Bertoncello2019,
  title = {Optimizing the Cell Culture Microenvironment},
  ISBN = {9781493990863},
  ISSN = {1940-6029},
  url = {http://dx.doi.org/10.1007/978-1-4939-9086-3_2},
  DOI = {10.1007/978-1-4939-9086-3_2},
  booktitle = {Mouse Cell Culture},
  publisher = {Springer New York},
  author = {Bertoncello,  Ivan},
  year = {2019},
  pages = {23–30}
}

@inbook{Poellen2023,
  title = {Cell Stretcher Assay to Analyze Mechanoresponses to Cyclic Stretching},
  ISBN = {9781071628515},
  ISSN = {1940-6029},
  url = {http://dx.doi.org/10.1007/978-1-0716-2851-5_6},
  DOI = {10.1007/978-1-0716-2851-5_6},
  booktitle = {Mechanobiology},
  publisher = {Springer US},
  author = {P\"{u}llen,  Robin and Konrad,  Jens and Hoffmann,  Bernd and Merkel,  Rudolf},
  year = {2023},
  pages = {91–105}
}

@article{Lax1960,
  title = {Fluctuations from the Nonequilibrium Steady State},
  volume = {32},
  ISSN = {0034-6861},
  url = {http://dx.doi.org/10.1103/RevModPhys.32.25},
  DOI = {10.1103/revmodphys.32.25},
  number = {1},
  journal = {Reviews of Modern Physics},
  publisher = {American Physical Society (APS)},
  author = {Lax,  Melvin},
  year = {1960},
  month = Jan,
  pages = {25–64}
}

@article{Godrche2018,
  title = {Characterising the nonequilibrium stationary states of Ornstein–Uhlenbeck processes},
  volume = {52},
  ISSN = {1751-8121},
  url = {http://dx.doi.org/10.1088/1751-8121/aaf190},
  DOI = {10.1088/1751-8121/aaf190},
  number = {3},
  journal = {Journal of Physics A: Mathematical and Theoretical},
  publisher = {IOP Publishing},
  author = {Godrèche,  Claude and Luck,  Jean-Marc},
  year = {2018},
  month = Dec,
  pages = {035002}
}
\end{document}